\documentclass[11pt]{article}
\usepackage{amsmath}
\usepackage{amsthm}
\usepackage{amssymb}

\def\half{\frac{1}{2}}

\def\pl{ \: + }

\newfont{\bbbold}{msbm10 scaled \magstep1}

\def\cM{{\cal M}}

\newfont{\goth}{eufm10 scaled \magstep1}

\def\gg{\mbox{\goth g}}

\def\a{\alpha}
\def\b{\beta}
\def\c{\gamma}\def\C{\Gamma}
\def\d{\delta}
\def\e{\epsilon}

\def\i{\iota}

\def\s{\sigma}

\def\th{\theta}

\def\be{\begin{equation}}\def\ee{\end{equation}}
\def\bea{\begin{eqnarray}}\def\eea{\end{eqnarray}}
\def\barr{\begin{array}}\def\earr{\end{array}}

\def\o{\omega}\def\O{\Omega}
\def\del{\partial}

\def\xz{\times}

\def\nab{\nabla}

\def\tN{\tilde{N}}

\def\hv{\widehat{v}}

\def\hL{{\widehat{L}}}\def\hN{\widehat{N}}

\def\pl{{(+)}} 
\def\Lf{I_{(4)}}
\def\Qf{J_{(4)}}
\def\Pf{K_{(4)}}
\def\Rs{I_{(6)}}
\def\Ie{I_{(8)}}
\def\Je{J_{(8)}}
\def\Ke{K_{(8)}}

\let\la=\label

\def\nn{\nonumber}
\def\bd{\begin{document}}
\def\ed{\end{document}}
\def\ba{\begin{array}}
\def\ea{\end{array}}
\def\bea{\begin{eqnarray}}
\def\eea{\end{eqnarray}}
\def\ft#1#2{{\textstyle{{\scriptstyle #1}\over {\scriptstyle #2}}}}
\def\fft#1#2{{#1 \over #2}}
\newcommand{\eq}[1]{(\ref{#1})}
\newcommand{\w}[1]{\\[0.#1cm]}
\def\eqs#1#2{(\ref{#1}-\ref{#2})}
\def\det{{\rm det\,}}
\def\tr{{\rm tr}}
\newcommand{\hoch}[1]{$\, ^{#1}$}
\newcommand{\tamphys}{\it\small Center for Theoretical Physics,
Texas A\&M University, College Station, TX 77843, USA}
\newcommand{\kings}
{\it\small Department of Mathematics, King's College, London, UK}
\newcommand{\uu}
{\it\small Department of Theoretical Physics, Uppsala, Sweden}
\newcommand{\hip}
{\it\small HIP-Helsinki Institute of Physics, P.O. Box 64 FIN-00014
University of Helsinki, Suomi-Finland}
\newcommand{\stock}
{\it\small Department of Theoretical Physics, Stockholm, Sweden}
\makeatletter
\renewcommand\theequation{\thesection.\arabic{equation}}
\@addtoreset{equation}{section} \makeatother

\def\balpha{{\overline{\alpha}}}
\def\bbeta{{\overline{\beta}}}
\def\bgamma{{\overline{\gamma}}}
\def\bdelta{{\overline{\delta}}}
\def\bepsilon{{\overline{\epsilon}}}
\def\bvarepsilon{{\overline{\varepsilon}}}
\def\bzeta{{\overline{\zeta}}}
\def\bareta{{\overline{\eta}}}
\def\btheta{{\overline{\theta}}}
\def\bvartheta{{\overline{\vartheta}}}
\def\biota{{\overline{\iota}}}
\def\bkappa{{\overline{\kappa}}}
\def\blambda{{\overline{\lambda}}}
\def\bmu{{\overline{\mu}}}
\def\bnu{{\overline{\nu}}}
\def\bxi{{\overline{\xi}}}
\def\bpi{{\overline{\pi}}}
\def\brho{{\overline{\rho}}}
\def\bvarrho{{\overline{\varrho}}}
\def\bsigma{{\overline{\sigma}}}
\def\bvarsigma{{\overline{\varsigma}}}
\def\btau{{\overline{\tau}}}
\def\bphi{{\overline{\phi}}}
\def\bvarphi{{\overline{\varphi}}}
\def\bchi{{\overline{\chi}}}
\def\bpsi{{\overline{\psi}}}
\def\bomega{{\overline{\omega}}}

\def\bR {\bb{R}}
\def\bb#1{\hbox{\mybb#1}}
\font\mybb=msbm10 at 11pt
\def\bH {\bb{H}}

\newcommand{\auth}
{\large P.S. Howe, G. Papadopoulos and V. Stojevic}

\begin{document}

\hfill{KCL-MTH-10-4}

\vspace{20pt}

\begin{center}
{\Large{\bf Covariantly constant forms on torsionful geometries from world-sheet and spacetime perspectives}}
\vspace{30pt}

\auth

\vspace{15pt}

\begin{center}
{\it\small Department of Mathematics, King's College, London, UK}
\end{center}


\vspace{60pt}

{\bf Abstract}

\end{center}

The symmetries of two-dimensional supersymmetric sigma models on target spaces with covariantly
constant forms associated to special holonomy groups are analysed. It is shown that each pair of such forms
gives rise to a new one, called a Nijenhuis form, and that there may be further reductions of the structure group.
In many cases of interest there are also covariantly constant one-forms which also give rise to symmetries. These geometries
are of interest in the context of heterotic supergravity solutions and the associated reductions are studied from a spacetime
point of view via the Killing spinor equations.

\setcounter{tocdepth}{3}
\pagebreak \tableofcontents \setcounter{page}{1}

\section{Introduction}

It has been known for some time \cite{Odake:1988bh, Delius:1989fy, Howe:1991ic, Shatashvili:1994zw, Figueroa-O'Farrill:1996hm} that
covariantly constant forms on a manifold $\cM$ give rise to W-type symmetries in the context of two-dimensional
$(1,0)$ and $(1,1)$ supersymmetric sigma models with target space $\cM$. At the same time, such forms arise naturally in
heterotic string backgrounds that preserve some of the spacetime supersymmetry, because the gravitino Killing spinor equation (KSE) is a parallel transport equation with respect to a metric connection $\nab^{(+)}$ with skew-symmetric torsion $H$. Therefore, all the forms that are
constructed as bilinears of the solutions of the gravitino KSEs are also $\nab^{(+)}$-parallel, and in turn, they generate W-type
of symmetries  (\ref{2.10}) in the worldvolume theory. In this paper we examine the geometries which admit special holonomy forms,
but which also have torsion, and investigate under which circumstances additional parallel forms exist and whether the structure
group of $\cM$ is reduced further due to their presence.

Before going into the details, the following general comments are in order.  The worldsheet and target space viewpoints are closely
related, clearly, since imposing conformal invariance in the sigma model at the quantum level is what defines the stringy equations of
motion \cite{Callan:1985ia}. Nevertheless, different aspects of the analysis are more natural from one viewpoint than the other.  On the
sigma model side, the algebra of the $W$-symmetries is a powerful tool, whose structure and closure tells us a lot about the geometry, already at
the classical level. It is also of mathematical interest to work with sigma models on target spaces for which the conformal anomaly
does not vanish, the most obvious example being a K\"{a}hler manifold that is not Calabi-Yau. The foremost disadvantages are that
target space spinors are difficult to describe from the worldsheet perspective, and that the dilaton arises at the order $\alpha'$, and
is more difficult to study without going into the intricacies of quantisation.  Therefore, studying the amount of supersymmetry preserved by
a particular background is most easily done in terms of the KSEs and the field equations of  heterotic supergravity.



The target space $(M,g,H)$ of $(1,1)$ and $(1,0)$ sigma models is an $n$-dimensional Riemannian manifold $(M,g)$ together with a closed three-form $H$, with $H=db$ locally. There are two natural metric connections with torsion,

\begin{equation}
\C^{(\pm)}{}^j_{ik}:=\C_{ik}^{j}\pm \half H^j{}_{ik}\ , \label{2.5}
\end{equation}

where $\C$ is the Levi-Civita connection. The torsion tensors of the two connections are given by

\begin{equation}
T^{(\pm)}{}^i_{jk}=\pm H^i{}_{jk}\ . \label{2.6}
\end{equation}

Any covariantly constant $(l+1)$-form $\o_L$, which we shall also see in the guise of a covariantly constant vector-valued $l$-form, $L$, i.e. $\nab^{(+)}L=0$, defines
a current

\be
j_L=L_{i L} D_{+}X^{iL}\ ,
\ee

which is conserved when the equations of motion are satisfied, $D_{-}j_L=0$ (or $\del_{--} j_L=0$ for (1,0) susy). (See appendix for conventions). The corresponding symmetry transformation is

\begin{equation}
\d_L X^i=a_L L^i{}_L D_+ X^L \label{2.10}
\end{equation}

where the parameter $a_L$ has Lorentz weight $-l$, Grassmann parity $(-1)^l$ and is independent of the minus coordinates. In the $(1,1)$ case one can have  similar symmetries for forms which are covariantly constant with respect to $\nab^{(-)}$, while in the $(1,0)$ case the $L$-type symmetries are restricted to one sector. For $l\geq 2$ these symmetries are non-linear and the associated
symmetry algebras are classical $W$-algebras. These have been studied, mainly in the torsion-free setting, in references \cite{Odake:1988bh, Delius:1989fy, Howe:1991ic, Shatashvili:1994zw, Figueroa-O'Farrill:1996hm, deBoer:2005pt}.

For the (1,1) case in the absence of torsion the left and right forms are identified
and it is natural to consider  those forms which are
associated with irreducible holonomy, given by Berger's list. The
structure group is reduced from $SO(n)$ to $U(m)$ or $SU(m)$, for
$n=2m$, $Sp(k)$ or $Sp(k)\cdot Sp(1)$, for $n=4k$, or to $G_2$ or
$Spin(7)$ for two exceptional cases in $n=7,8$ respectively.

The key point is that, when $H$-flux is turned on, consideration of
the worldsheet W-algebra shows that  there may be additional
covariantly constant forms which are not simply functions of the
original set of special holonomy forms; when this is the case, there
will be further reductions of the structure group. Indeed,
associated with any pair $(L,M)$ of such forms there is a
covariantly constant form $\tN(L,M)$ which we call the Nijenhuis
form; it is related to the Nijenhuis concomitant but is not, in
general, the same object.

The simplest case, and perhaps most important from a physical point
of view, is when we have an almost  complex structure, $I$. If this
is integrable, there is a second supersymmetry, and if we have a
pair of them, $I^{(\pm)}$, we get a $(2, 2)$ supersymmetric sigma
model for which the target space is bi-Hermitian \cite{Gates:1984nk,Howe:1985pm}. Such geometries
are now known to be equivalent to generalised K\"ahler geometries \cite{Hitchin:2004ut,Gualtieri:2003dx, Lindstrom:2005zr, Hull:2008vw}, and
have been much studied in the recent literature \cite{Grana:2004bg, Grana:2005sn, Grana:2005jc, Grana:2005ny, Grana:2006hr, Stojevic:2008qy}. Here we consider
geometries for which the $I$s are not complex. This problem was
investigated some time ago in \cite{Delius:1989fy} where it was
shown that the Nijenhuis tensor $N$, which is equal to the Nijenhuis
form $\tN(I,I)$ in this case, is covariantly constant and hence
defines a new symmetry. Moreover, it appears that it generates an
infinity of higher-order symmetries. The question of further
reduction of the structure group was not addressed in
\cite{Delius:1989fy}, but this certainly does occur. It is not
difficult to show that $N$ is $(3,0)+(0,3)$ with respect to a
hermitian basis defined by $I$, so that, for example for $m=3$, the
structure group is automatically reduced to $SU(3)$.\footnote{An example of such a manifold is $S^6$ equipped with the standard almost complex structure, which reduces the structure to $U(3)$.  But since the almost
complex structure is not integrable, the Nijenhuis tensor does not vanish reducing the structure further to $SU(3)$.}

For (1,1) models the most general case would involve two independent sets of left and right
special holonomy forms which give  rise to two structure groups
$G^{(\pm)}$ which need not be isomorphic in principle. If they are,
but the left and right forms are not the same, we have what might be
called a bi-$G$-structure. We shall focus for the most part on one
sector and investigate which sort of reductions can arise. Studying both sectors
is relevant to type II  string theories. The bi-$G$-structures do not arise for the $(1,0)$ sigma model, which corresponds to the heterotic string.


From the point of view of heterotic supergravity additional invariant one-forms arise when the dilatino KSE is not satisfied. The KSEs have been solved in generality in
\cite{het1, het2}. The existence of $\nab^{(+)}$-parallel
spinors, and so solutions of the  heterotic string gravitino KSE,  requires that the holonomy of $\nab^{(+)}$ reduces
to a subgroup of the isotropy group $G$ of the parallel spinors in
$Spin(9,1)$.  These
isotropy groups are non-compact
\bea
&&Spin(7)\ltimes\bR^8(1)~,~~~SU(4)\ltimes
\bR^8(2)~,~~Sp(2)\ltimes\bR^8(3)~,~~~\times^2Sp(1)\ltimes\bR^8(4)~,~~~
\cr &&Sp(1)\ltimes \bR^8(5)~,~~~ U(1)\ltimes \bR^8(6)~,~~~\bR^8(8)~,
\la{noncomp}
\eea

and compact

\bea
G_2(2)~,~~~SU(3)(4)~,~~~SU(2)(8)~,~~~\{1\}(16)~, \la{ns1}
\eea

where in parenthesis is the number of invariant spinors.
So, the solution of the gravitino
KSE leads to the investigation of manifolds with
$G$ structures equipped with a compatible connection with
skew-symmetric torsion.

Apart from the parallel forms that can be constructed as parallel
spinor bilinears, the spacetime,  under certain conditions,  may
admit additional $\nab^{(+)}$-parallel forms. The presence of these
forms leads to a  further reduction of the holonomy of $\nab^{(+)}$
to a subgroup of $G$. Such additional forms have been found in
\cite{het2} by analyzing the  dilatino KSE. In
particular, the conditions that arise in the dilatino KSEs can be stated as the vanishing of certain forms, which we refer to as $\tau$-forms. Now if
the dilatino KSE is not satisfied, the $\tau$-forms are no longer vanishing. However, it can be shown to be
$\nab^{(+)}$-parallel subject to enforcing a Bianchi identity,
$dH=0$ and the field equations of the theory. On one hand the existence
of these forms breaks supersymmetry because the dilatino KSE is not satisfied, but on the other it leads to structure group reduction and thus the existence of additional parallel spinors. There are many examples of backgrounds which are of this type. As example one can take the WZW models with constant dilaton. These are non-supersymmetric backgrounds, have 16 parallel spinors in the context of heterotic supergravity, and solve the field equations.


The backgrounds that solve the gravitino KSE are
Lorentzian. To adopt the  analysis to the case of sigma models for
which the target space is Euclidean, we shall extract the
``Euclidean component'' of the Lorentzian supersymmetric
backgrounds. To do this,
we make some simplifying assumptions on the
structure of the Lorentzian manifolds. These assumptions are
dictated by the geometry of the Lorentzian supersymmetric
backgrounds  and depend on whether the isotropy group $G$ of the
parallel spinors is non-compact or compact. A more detail explanation will be given
in sections \ref{sec:target_compact} and \ref{sec:target_noncompact}.
Typically, we shall assume
that the spacetime is metrically $\bR^{9-n,1}\times X_n$, the fields are
independent from the $\bR^{9-n,1}$ coordinates, and the 3-form flux has components
only along $X_n$, where $X_n$ is the Euclidean component of the spacetime.
In such a case, the holonomy of $\nab^{(+)}$ reduces to a subgroup of
\bea
&&Spin(7)[8]~,~~~SU(4)[8]~,~~~\times^2Sp(1)[8]~,~~~
Sp(1)[8]~,~~~ U(1)[8]~,~~~
\cr
&&G_2[7]~,~~~SU(3)[6]~,~~~SU(2)[4]~,  ~~~Sp(2)[8]~
\la{eucomp}
\eea
where the number in the square brackets is the dimension $n$ of $X_n$.

All the $X_n$ manifolds admit $\nab^{(+)}$-parallel forms which are the fundamental forms of the groups (\ref{eucomp}).
As we have mentioned, these give rise to W-symmetries of the string world-volume action.
We shall show that if the fields  do not satisfy the dilatino KSE,
then $X_n$ admits additional  $\nab^{(+)}$-parallel forms
subject to a Bianchi identity,  $dH=0$ and the field equations of the heterotic supergravity.
As a result, the string world-volume theory in these
backgrounds admits additional symmetries, and the holonomy group of $\nab^{(+)}$ on $X_n$
reduces to proper subgroup of those of (\ref{eucomp}). We shall investigate the patterns of
reductions in each case. We shall demonstrate
that in many occasions, the gravitino KSE admits additional parallel spinors which in turn
trigger further reduction the holonomy.

In section \ref{sec:comm_algebra} we discuss the algebra of $L$-type symmetries due to special holonomy forms in a general setting. In section \ref{sec_nijenhuis_forms}  we examine how new forms are generated both from the  W-algebra and heterotic supergravity perspective. In section  \ref{subsec:worldsheet_algebras} we examine the W-algebras in detail going through the special holonomy list, and in section  \ref{sec:sugra} we go through the same list from the heterotic supergravity perspective and also examine the structure group reduction in detail. We give some concluding remarks in \ref{sec:conclusion}. The appendix summarises our notation and conventions.

\section{Commutator algebra}
\label{sec:comm_algebra}

In this section we compute the commutator of two symmetry transformations of the type given in \eq{2.10}, focusing for the most part on transformations of the same chirality. We shall deal with left symmetries (for which the parameters depend only on the plus coordinates) but there will be no need to litter the formulae with pluses on the $L$-tensors. The general expression for the commutator of symmetries based on special holonomy forms was computed in \cite{Howe:1991ic}; here we shall rewrite this so that we can identify the symmetries that arise in the presence of torsion. This was briefly outlined in \cite{Howe:2005je}.  Understanding the algebra of symmetries not stemming just from the special holonomy forms, but including for example the Nijenhuis or $\tau$-forms, is best considered case by case.  We delay the discussion of this to sections \ref{sec_nijenhuis_forms} and \ref{subsec:worldsheet_algebras}.


The commutator is

\begin{equation}
[\d_L,\d_M]X^i =\d^{(1)}_{LM}X^i + \d^{(2)}_{LM}X^i +
\d^{(3)}_{LM}X^i\ , \label{3.1}
\end{equation}

where

\begin{eqnarray}
\d^{(1)}_{LM}X^i &=& a_M a_L N(L,M)^i{}_{LM} D_+X^{LM}\ , \la{3.2}\w1
\d^{(2)}_{LM}X^i &=& \left(-m a_M D_+ a_L (L\cdot M)_{jL_2,iM_2}
+l(-1)^{(l+1)(m+1)}a_L D_+ a_M (L\cdot M)_{iL_2,jM_2}\right) \xz \nn \w1
&\phantom{=}& \xz\, D_+ X^{jL_2M_2}\ ,\la{3.3} \w1
\d^{(3)}_{LM}X^i &=& -2ilm(-1)^l a_M a_L (L\cdot
M)_{(i|L_2|,j)M_2} \del_{++} X^j D_+ X^{L_2 M_2}\ .
\la{3.4}
\end{eqnarray}

Here

\begin{equation}\label{3.5}
(L\cdot M)_{iL_2,jM_2}:=L_{ki[L_2} M^k{}_{|j|M_2]}\ ,
\end{equation}

while $N(L,M)$ denotes the Nijenhuis concomitant of $L$ and $M$. We recall that a vector-valued $l$-form $L$ defines a derivation, $\i_L$, of degree $l-1$ of the algebra of differential forms (i.e. a linear map sending $p$-forms (in $\O_p$) to $p+l-1$-forms which satisfies the graded Leibniz rule) by

\begin{equation}\label{3.6}
\O_p \ni \o \mapsto \i_L\o := p L^i{}_L \o_{i P_2} dx^{LP_2} \in
\O_{p+l-1}\ .
\end{equation}

Since the graded commutator of two derivations is also a derivation we can define a new derivation $d_L$ by

\begin{equation}\label{3.7}
 d_L := \i_L d +(-1)^l d \i_L \ .
\end{equation}

Clearly $\i_L$ generalises the interior product of a form with a vector field $v$ while $d_L$ generalises the Lie derivative along $v$. Given two vector-valued forms we then find that

\begin{equation}\label{3.8}
[d_L,d_M]= d_L d_M -(-1)^{lm} d_M d_L =d_{N(L,M)}\ .
\end{equation}

This equation defines the Nijenhuis concomitant $N(L,M)$. When $L=M=I$, an almost complex structure, $N(I,I)$ is the usual Nijenhuis tensor. The explicit formula is

\begin{equation}\label{3.9}
N(L,M)^i=\left(L^j{}_L \del_j M^i{}_M -M^j{}_M \del_j L^i{}_L
-lL^i{}_{jL_2} \del_{l_1} M^j{}_M +m M^i{}_{j M_2} \del_{m_1}
L^j{}_L\right) dx^{LM} \ .
\end{equation}

In this formula we can replace the ordinary derivatives by the Levi-Civita covariant derivative. For torsion-free sigma models on special holonomy target spaces, therefore, the Nijenhuis concomitants vanish.

There are at most three independent symmetries in the commutator but they do not correspond directly to the division exhibited in \eq{3.1}. To elucidate this structure we begin by writing the Nijenhuis term in terms of the torsion, making use of the fact that both $L$ and $M$ are covariantly constant with respect to $\nab^\pl$. One finds

\begin{eqnarray}
N(L,M)^i{}_{LM}&=& -H^i_{jk} L^j{}_L M^k{}_M +mL^j{}_L H_{jm_1}{}^k
M^i{}_{kM_2} -lM^j{}_M H_{jl_1}{}^k L^i{}_{kL_2}+\nn\w1
&\phantom{=}& +  lm \, H_{k l_1 m_1}(L\cdot
M)^{(i}{}_{L_2},{}^{k)}{}_{M_2}\ .
\label{3.10}
\end{eqnarray}

The first line of the right-hand side is totally antisymmetric (when the $i$ index is lowered), but is not covariantly constant in general. However, one can always add to it a term so that a covariantly constant $(l+m+1)$-form results. In order to see this and to simplify the remaining terms we use the following algebraic results, which can be proved for any of the special holonomy forms,

\begin{eqnarray}
\label{eq:spechol_relations}
(L\cdot M)_{i[L_2,jM_2]} &=& (-1)^{l+1}P_{ijL_2 M_2} +\frac{m}{2}g_{i[j}Q_{L_2
M_2]}\ ,\nn\w1
(L\cdot M)_{[jL_2,|i|M_2]} &=& (-1)^{l}P_{ijL_2 M_2} +\frac{l}{2}g_{i[j}Q_{L_2
M_2]}\ ,\nn\w1
(L\cdot M)_{i[L_2,|j| M_2]} + (i\leftrightarrow j)&=& g_{ij} Q_{L_2
M_2}-(l+m-2) g_{(i[l_2} Q_{j)L_3 M_2]}\ .
\la{3.11}
\end{eqnarray}

The tensors $P$ and $Q$ are functions of the special holonomy forms and the metric that can be found from the above equations; they are totally antisymmetric and covariantly constant; in particular cases they can vanish.\footnote{A simple example is  $L=M=I$, an almost complex structure, in which case $P=0$ and $Q=1$.} Both of them can be used to define $L$-type symmetry transformations, but in the commutator $[\d_L,\d_M]$ $Q$ is combined with the energy-momentum tensor. After some algebra one then finds that

\begin{equation}
[\d_L,\d_M]X^i =\d_P X^i + \d_{\tN} X^i + \d_K X^i\ , \label{3.12}
\end{equation}

where each term is now a symmetry by itself. The $P$ transformation, which is of standard $L$-type has parameter $a_P$ given by

\be
a_P=(-1)^{l+1} m a_M D a_L -(-1)^m l D a_M a_L \ .
\la{3.13}
\ee

The $\tN$ transformation is also of this type; The Nijenhuis form $\tN$ is given by

\be
\tN_{iLM}=-(l+m+1)\left( H_{jk[i} L^j{}_{L} M^k{}_{M]}+
(-1)^{l}\,\frac{lm}{6} H_{[il_1 l_2} Q_{L_3 M]}\right)\ .
\la{3.14}
\ee

It is not the Nijenhuis concomitant, since the latter is not totally antisymmetric in general, but it is constructed from it. It is not immediately obvious that $\tN$ is covariantly constant. The proof of this fact, together with a discussion of these forms for the various special holonomy groups, is given in a separate section. The parameter $a_{\tN}$ is just $a_M a_L$.

Finally, we consider the $K$ transformation. If we define

\be
K_{i,K}:=g_{i[k_1} Q_{K_2]}\ ,
\la{3.18}
\ee

where the multi-index $K$ takes on $l+m-1$ values, then it is not difficult to show (for any covariantly constant antisymmetric tensor $Q$) that

\bea
\d_K X_i&=&\frac{l+m-1}{l+m-2}\Big(
a_K K_{j,iK_2} \del_{++} X^j D_+ X^{K_2}+\frac{i(-1)^k}{k}
\nab^\pl_+ (a_K D_+ X^K) \nn \w1
&\phantom{=}& + \frac{ia_K}{k}(H_{ijk_1}
Q_{K_2}-\frac{k+2}{6}H_{[ijk_1} Q_{K_2]})D_+ X^{jK}\Big)
\la{3.19}
\eea

is a symmetry of the action \eq{2.1}. In fact, the corresponding conserved quantity is the composite current $T Q$. For the case in hand the parameter $a_K$ is

\be
a_K=i(-1)^{l+1}\frac{lm(l+m-2) }{2}a_M a_L\ .
\la{3.20}
\ee

In summary, the commutator of two symmetries  within the same sector determined by covariantly constant special holonomy forms generically gives rise to three symmetries, two of which again involve covariantly constant forms and the third being generated by the product of the energy-momentum and the current of another form. For $(1,1)$ sigma models the commutator of a left and a right symmetry closes up to equation of motion terms \cite{Howe:2005je}.

\section{New invariant forms in the presence of $H$-flux}
\label{sec_nijenhuis_forms}

In this section we describe the invariant forms that are potentially  generated by the algebra of $L$-type symmetries when $H$ is non-zero, as well as a set of invariant one-forms  that arise from the target space perspective when the dilatino KSE is not satisfied. We refer to the former as Nijenhuis forms, which we denote as $\tN$, and the latter as $\tau$-forms. The $\tau$ forms are not generated by the special holonomy algebras and need to be introduced into the $\sigma$-model as additional symmetries.


For any two special holonomy forms, $L,M$, we define the Nijenhuis form $\tN(L,M)$ by

\be
\tN_{iLM}=-(l+m+1)\left( H_{jk[i} L^j{}_{L} M^k{}_{M]}+
(-1)^{l}\,\frac{lm}{6} H_{[il_1 l_2} Q_{L_3 M]}\right)\ .
\la{3.23}
\ee

This is covariantly constant provided that the algebraic relations (\ref{3.11}) are satisfied.  With $H$ turned on the algebra in general generates new forms and these in general obey relations different to those in (\ref{3.11}).

To prove the covariant constancy of (\ref{3.23}) one uses the Bianchi Identity, together with the fact that $H$ is closed, to obtain

\be
\nab^\pl_p H_{ijk}= 3R^\pl_{[ij,k]p}\ .
\la{3.24}
\ee

We use the convention that the form indices on the curvature are the second pair. Applying $\nab^\pl_p$ to the first term in $\tN$ we find

\be
\nab^\pl_p H_{jk[i}
L^j{}_{L}M^k{}_{M]}=(R^\pl_{[i|j,kp|}+R^\pl_{k[i,|jp|}+R^\pl_{jk,[i|p|})L^j{}_{L}
M^k{}_{M]} \ .
\la{3.25}
\ee

The first two terms on the right vanish because $L$ and $M$ are invariant tensors under the holonomy group and the first pair of indices on the curvature take their values in the corresponding Lie algebra. Using the same fact, we see that the third term can be written

\bea
R^\pl_{jk,[i|p|}L^j{}_{L} M^k{}_{M]}&=&l R^\pl_{j[l_1,i|p|})L^j{}_{kL_2}
M^k{}_{M]}\nn\w1
&=&-l R^\pl_{j[l_1,i|p|}(L\cdot M)^j{}_{L_2,M]}\nn\w1
&=&\frac{lm}{2}(-1)^{l+1} R^\pl_{[l_1 l_2,i|p|}Q_{L_3 M]}\ ,
\la{3.26}
\eea

where, in the last line, we have used \eq{3.11} and the invariance of $P$. From this it is easy to see that $\tN$ is covariantly constant as claimed.

A further complication is that, as we shall discuss concretely below, some of the $\tN$ forms can induce a split in the tangent space, i.e. a reduction of the structure group to a product of two smaller groups. This  implies that a covariantly constant almost-product structure $\mathcal{R}^{i}_{ \ j}$ is present:
\begin{equation}
\label{eq:product_structure}
\mathcal{R}^2 = 1 \ \ \ , \ \ \  \nabla^{(+)}_k \mathcal{R}^{i}_{ \ j} = 0  \ .
\end{equation}
In general these structures are not integrable. Symmetries of $(1,1)$ models associated with covariantly constant almost-product structures have been studied in detail in \cite{Stojevic:2009ub}. Integrability is equivalently expressed as the vanishing of the mixed parts of $H$ with respect to the projectors

\begin{equation}
\label{eq:projectors}
\mathcal{P}^i_{ \ j}  :=  \frac{1}{2} \left(  \delta^i_{ j} + \mathcal{R}^i_{ \ j} \right)  \ \ \  ,  \ \ \   \mathcal{Q}^i_{ \ j} := \frac{1}{2} \left( \delta^i_j - \mathcal{R}^i_{ \ j} \right) \    .
\end{equation}

Without assuming integrability one needs to carefully work out the combined algebra of the superconformal symmetries associated with non-integrable projectors together with $L$-type symmetries. The complication is compounded by the fact  that the projected version of the superconformal transformation contains a non-linear piece involving the mixed part of $H$, and the effect of this non-linearity needs to be carefully considered. Furthermore, without going through this analysis, we do not know the appropriate generalisation of (\ref{3.11}).  In this paper we will not attempt to understand the general algebra, but will nevertheless be able to get a handle on many of the lower dimensional cases, because the analysis reduces to studying symmetries related to covariantly constant one-forms.


For this reason, and also in order to incorporate symmetries associated with the $\tau$-forms (we discuss these shortly), it will be useful to spell out the symmetry algebra involving a $\nabla^{(+)}$-invariant vector $v$.

The symmetry transformation is given by:

\begin{equation}
\label{eq:v_symmetry}
\delta X^i = a_v v^i \ ,
\end{equation}

and has the associated conserved current $v_i D_+ X^i$. The commutator with a superconformal symmetry closes again to (\ref{eq:v_symmetry}),    and the commutator with an $L$-type symmetry yields two further $L$-type symmetries. The first of these is due to the $l$-form

\begin{equation}
\label{eq:Lv_form}
i_v \omega_L  \ ,
\end{equation}

while the second is due to the $(l+1)$ Nijenhuis form

\begin{equation}
\label{eq:vv_form}
\tN(v, L)_{i_1 i_2 \cdots i_{l+1}} = v^m H_{m p  [ i_1} L^p_{ \ i_2 \cdots i_{l+1} ] }  \ .
\end{equation}

Clearly the former does not reduce the structure group, but the latter potentially does.

The commutator of two vector-type symmetries (\ref{eq:v_symmetry}) associated with  $v$ and $w$ yields a Nijenhuis one-form. This object is essentially the Lie bracket with the index lowered, and can be written as

\begin{equation}
\label{eq:vv_form2}
\tN(v, w)_i = H_{ i jk} v^j w^k \ ,
\end{equation}

using the covariant constancy of $v$ and $w$. It follows that the structure group is potentially reduced further if $v$ and $w$ are linearly independent.

The \emph{Lee form} of a general form $\omega_L$ is defined as

\bea
\theta_{L}=- k_L \star(\star d\omega_L  \wedge \omega_L) \ .
\eea

The constants $k_L$ are determined by the requiring that the $\tau_{\omega_L }$ one-form, defined as

\begin{equation}
\label{eq:tau_def}
\tau_{ L } := \theta_{L}-2d\Phi \ ,
\end{equation}

is covariantly constant \cite{het2} when we use the equations of motion of the heterotic string to the lowest order in $\alpha'$, with $\Phi$ as the dilaton field and the gauge fields set to zero. For the particular examples we consider, the constants $k_L$ are all listed in section \ref{sec:sugra}.


The equations of motion coming from the metric and $b$-field $\beta$-functions are :

\begin{align}
\label{eq:stringy_eqs1}
&  R_{ij} - \frac{1}{4} H_{ikl} H_{j}^{ \ kl} + 2 \nabla_{ i } \nabla_{ j} \Phi = 0 \ ,  \\ \nonumber
&  \nabla^k H_{kij} - 2 (\nabla_k \Phi) H^k_{ \ ij} = 0  \ .
\end{align}

The equation of motion coming from the dilaton $\beta$-function is:

\begin{align}
\label{eq:stringy_eqs2}
4 ( \nabla \Phi )^2 - 4 \nabla^2 \Phi - R + \frac{1}{12} H_{ijk}  H^{ijk} + \frac{(D - 10)}{3 \alpha'} = 0 \ .
\end{align}

It can be seen by contracting the first equation in (\ref{eq:stringy_eqs1}) that  for a constant dilaton, and when $D=10$, the equations of motion can only be satisfied if also $H=0$.

Equations (\ref{eq:stringy_eqs1}) are conveniently expressed in terms of  $R^{(+)}_{ij}$,  the Ricci tensor of the $\nabla^{(+)}$ connection, as:

\begin{equation}
\label{eq:stringy_eom_alt}
R^{(+)}_{ij}-2\nab^{(+)}_j \nab_i \Phi =0 \ .
\end{equation}

For all the examples we consider the Lee form of $\omega_L$ turns out to be the contraction of some covariantly constant 4-form $\lambda$ and $H$:

\begin{equation}
(\theta_{ L } )_i  \propto   \lambda_{i jkl} H^{jkl} \ .
\end{equation}

The covariant  constancy of $\tau_{L}$ can be demonstrated straightforwardly from  (\ref{3.24}) and (\ref{eq:stringy_eom_alt}).  For example, for the Lee form $\theta_I$  associated with an almost-complex structure, we have

\begin{equation}
(\theta_I)_m  \propto I_{[ ij} I_{km ] } H^{ijk} \  .
\end{equation}

It follows from (\ref{3.24}) that $\nabla^{(+)} \theta_ I$ is proportional to $R^{(+)}_{ij}$, provided that also

\begin{equation}
I^{ij} R^{(+)}_{ij k m} = 0 \ ,
\end{equation}

which is the requirement for the structure group to be in $SU(m)$, rather than just $U(m)$. It is then obvious from (\ref{eq:stringy_eom_alt}) that  $ \nabla^{(+)} \tau_I = 0$ imposes the metric and $b$-field equations of motion.

\section{Worldsheet symmetry algebras}
\label{subsec:worldsheet_algebras}

In this section we discuss how the special holonomy algebras are deformed in the presence of $H$-flux by the Nijenhuis and $\tau$ forms. The full analysis of the structure group reduction is left to section \ref{sec:sugra}.


\subsection{$G=U(m)$ and $G=SU(m) ;\ n=2m$}
\label{subsec:Um}


The reduction of the holonomy group to $U(m)$ is associated with an
almost complex structure $I$. In this case,  which has been studied
for $m=3$ in \cite{Delius:1989fy}, one is dealing with the usual
almost-complex Nijenhuis form. In a hermitian frame basis the torsion
can be decomposed into $(3,0)$ and $(2,1)$ components, together with
their complex conjugates. It is easy to see that the Nijenhuis form
is proportional to the $(3,0)$ plus $(0,3)$ part of $H$. Combined with (\ref{3.24}) this
provides another way of seeing that it is covariantly constant as
the curvature tensor is pure on its Lie algebra indices, and
therefore mixed when one of them is lowered. Although the Nijenhuis
form is identically covariantly constant, it still implies a further
reduction for the structure group.

Further invariant forms arise if, in addition to $I$, we have a non-integrable almost-product structure $\mathcal{R}$ (\ref{eq:product_structure}) covariantly constant with respect to $\nabla^{(+)}$. In \cite{Stojevic:2009ub} it is shown that, if we let $\{ a, b, c \} $ and  $ \{ a', b', c' \}$ denote indices associated with subspaces projected onto by $\mathcal{P}$ and $\mathcal{Q}$, then in addition to the purely (anti)-holomorphic components of $H$,  which are related to the almost-complex Nijenhuis form, also

\begin{equation}
\label{eq:Hmixed_cov_constant}
H_{a b \overline{c'} } \ \ \ ,    \ \ \ H_{ a' b' \overline{c}} \ ,
\end{equation}

as well as the complex conjugate components, are covariantly constant.


In addition to an almost complex structure, the $SU(m)$ holonomy
group is associated with two real $m$-forms,  $L$ and $\hL$ which
are the real and imaginary parts of a complex $(m,0)$ form (with
respect to a hermitian frame). In this case we have a number of
possible Nijenhuis forms, $\tN(I,L),\tN(I,\hL),
\tN(L,L),\tN(\hL,\hL),\tN(L,\hL)$ as well as $\tN(I,I)$ which we
will write as $N$. Apart from $N$ it turns out that the only
non-vanishing ones are $\tN(I,L)$ and $\tN(I,\hL)$ and these are
given in terms of $N$ and $L,\hL$. Therefore the only further
reduction of the structure group is due to the presence of the
$(3,0)+(0,3)$ form $N$.

As an example we sketch the proof that $\tN(L,L)$ vanishes for $m$
even (it is identically zero for $m$ odd).  This is a $(2m-1)$-form
so that it is convenient to look at its one-form dual. The
$(2m-4)$-form $Q$ is proportional to $I^{m-2}$ in this case so that
the dual of the second term in $\tN$ (equation \eq{3.23}) is
proportional to $I^2_{ijkl} H^{jkl}$. To evaluate the dual of the
first term we use the fact that $L$ is self-dual for $m$ even to
arrive at an expression of the form $L_{ij}^{ \ \ p_1\ldots p_{m-2}}
L_{kl p_1\ldots p_{m-2}}H^{jkl}\propto I^2_{ijkl} H^{jkl}$. A
careful evaluation shows that the two terms cancel. The vanishing of $\tN(\hL,\hL)$ and $\tN(L,\hL)$ can be
verified in a similar fashion.


Now consider $\tN(I,L)$. In this case $Q=0$ since it involves the
double contraction of $I$ and $L$ which are  of different type with
respect to the almost complex structure. It is again easier to look
at the dual, which is an $l$-form in this case (recall that
$m=l+1$). We find, for $m$ even,

\be
*\tN_{i_1 \ldots
i_l}=\frac{1}{m}\left(lH_{[i_1}{}^{jk}\hL_{j k i_2\ldots i_l]} + L_{k
i_1\ldots i_l} I^{pq} H_{pq}{}^k\right)\ .
\la{3.27}
\ee

Because $L$ and $\hL$ are both of type $(m,0)+(0,m)$ it follows that
this expression can be either $(l,0)$ or $(l-1,1)$ or  complex
conjugates. It is not difficult to verify that the $(l,0)$ part
vanishes and this implies that only the $(3,0)+(0,3)$ components of
$H$ contribute. But this is just $N$, so we find

\be
*\tN_{1_1\ldots i_l}=\frac {l}{4(l+1)}N_{[i_1}{}^{jk}\hL_{j k i_2\ldots
i_l]}\ .
\la{3.28}
\ee

Similar expressions can be derived for $m$ odd and for $\tN(I,\hL)$.
These forms, although non-zero, are generated from the  original set
together with $N$ so that there is no further reduction of the
structure group.

Next we need to consider Nijenhuis forms involving $N$. For $m>3$ $N$ will induce a split in the tangent space. Due to the complications which were summarised in the context of (\ref{eq:product_structure}), the analysis in the next paragraphs applies only when the almost product structure associated with this split is integrable.

$\tN(I,N)_{ijkl} \propto H_{rs [ i} I^r_{ \ j} N^s_{ \ kl]}$ is potentially
non-vanishing.  By going to a hermitian frame one can see that the
$(3,1)$ and $(1,3)$ parts vanish while the $(2,2)$ part involves
only components of $N$. The only part we need consider is therefore
the $(4,0)$ part. It is not difficult to see that it is proportional
to $(H^m{}_{[ij}H_{kl]m})_{4,0}$. On the other hand $dH=0$ can be
expressed as

\begin{equation}
\label{eq:dH_constraint}
(dH)_{ijkl} = \nabla^{(+)}_{[i} H_{jkl]} + \frac{3}{2}H^m_{ \ [ij} H_{kl]m} =0 \ .
\end{equation}

In projecting out the  $(4,0)$ component we eliminate the first
term, since it is simply the covariant derivative of $N$,  and the
remainder of the expression implies that the $(4,0)$ component of
$\tN(I,N)$ vanishes.  Also, from the $(2,2)$ part of
(\ref{eq:dH_constraint}) we can see that the condition $dH=0$ is
incompatible with setting $H=N$, because then the $(2,2)$ component of
(\ref{eq:dH_constraint}) would be inconsistent unless $H$ itself
vanishes.

We also need to consider the five-form $\tN(N, \hN)$, where $\hN_{jlm}$ is given
by $I^{k}_{ \ [ j} N_{lm]k}$ ($\tN(N, N)$ vanishes identically).


If the almost-product structure (\ref{eq:product_structure}) is integrable, it follows straightforwardly that the $(5,0)+(0,5)$  and $(4,1) + (1,4)$ components of $\tN(N, \hN)$
are constructed from known covariantly constant tensors, and thus
have no impact on the structure group. The former is zero for $m=3$ and $m=4$, and the latter is non-zero for $m>3$. On the other hand, the $(3,2) +(2,3)$ component involves mixed parts of torsion and can potentially reduce the structure group. However, it turns out that at least for the $m=3,4$ cases such contributions vanish.


When we are in $SU(m)$ rather than $U(m)$, taking $\theta_{I}$ to be covariantly constant with respect to $\nabla^{(+)}$ implies that $R^{(+)}_{ij} = 0$. As discussed at the end of the previous section, these are the stringy equations of motion up to dilaton terms. Since $\theta_{I}$ contracted with $I$ is proportional to the trace of $H$,  the vanishing of $\theta_{I}$ is equivalent to the primitivity condition $I^{ij}H_{ijk} = 0$ which often arises in the literature. The one-form

\begin{align}
\label{eq:mod_primitivity}
I^{jk} H_{ijk} - I_i{}^k \nab_k \Phi \
\end{align}

is  proportional to $\tau_{I}$ contracted with $I$, and is  covariantly constant under $\nabla^{(+)}$ provided that the stringy equations of motion (\ref{eq:stringy_eom_alt}) are satisfied, which is what we assume in the rest of this section.  It is not difficult to show that  $\theta_{L}$ is equal to $\theta_{I}$, so we potentially only have a single additional symmetry due to the $\tau$ one-forms.



We now consider the particulars of the $m=3,4$ cases without restricting any almost-product structures that arise in the analysis to be integrable.

For $m=3$ it is obvious that the structure group is reduced to $SU(3)$. Since $N$ and $\hN$ are respectively the  real and complex parts of the holomorphic volume form,  we have $\tN(N, \hN)=0$, and the algebra closes (as a W-algebra). The $(4,0)+(0,4)$ part of
(\ref{eq:dH_constraint}) is now trivially zero. If $H$ is primitive the algebraic part of the $(3,1)+(1,3)$ component of
(\ref{eq:dH_constraint}) is zero, and following from  (\ref{3.24}) the differential part implies relations between components of the
curvature tensor.

The symmetry due to $\tau_{I}$  is not generated from the original $SU(3)$ special holonomy symmetries, and must therefore be introduced separately. Let us introduce the hermitian basis of one-forms as $e^\alpha = \{ e^a, e^3 \}$, $a = \{ 1,2 \}$,  together with their complex conjugates, which we choose so that  $\tau_{I}$ is entirely in the $e^3$/$e^{\overline{3}}$ directions. This reduces the structure group to $SU(2)$.

It follows from the general result (\ref{eq:Hmixed_cov_constant}) that,

\begin{equation}
\label{eq:m3_cov_const_comp}
H_{ab \overline{3}} \ \ \ ,  \ \ \ H_{\overline{a} \overline{b} 3}  \ \ \ , \ \ \  H_{ a \overline{3} 3}  \  \ \ , \ \ \  H_{ \overline{a} 3  \overline{3} } \ ,
\end{equation}

are all covariantly constant.  The presence of the latter two components reduces the the structure group further to the trivial group. The covariant constancy of the components (\ref{eq:m3_cov_const_comp}) can also be inferred by considering $\tN(I, \tau_{I})$, and the commutator between symmetries associated  with $\tau_I$ and  $\overline{\tau}_I$.

For $m =  4$ we introduce a hermitian basis of forms $e^{\a},\ \a=1,\ldots 4$,  together with
their complex conjugates $e^{\bar\a}$. The covariant constancy of $N$ implies that we can split an hermitian basis of one-forms as
$e^\alpha = \{ e^a, e^4 \}$, $a = \{ 1,2,3 \}$, such that

\be
\label{eq:m4}
N_{abc}=\e_{abc};\qquad N_{ab4}=0\ .
\ee

Clearly the structure group is reduced to $SU(3)\xz U(1)$.  In $SU(4)$ there are also covariantly constant $(4,0)$ and $(0,4)$  forms and so the  structure group reduces to
$SU(3)$. In this case we can study the symmetry algebra generated by the dual one-forms of $N$ and $\hN$ which we call $v$ and $\hv$.    It follows from   (\ref{eq:vv_form2}) that the components

\begin{equation}
\label{eq:m4_reduction1}
H_{a 4 \overline{4}}   \ \ \ , \ \ \   H_{\overline{a} 4 \overline{4}}   \ ,
\end{equation}

are covariantly constant. Their presence reduces the structure group to $SU(2)$. Furthermore, it follows from  the covariant constancy of $\tN(I, v)$ that

\begin{equation}
\label{eq:m4_reduction2}
H_{ab 4} \ \ \ ,  \ \ \ H_{ ab \overline{4}}
\end{equation}

are covariantly constant (actually, $H_{ab4}=0$ due to the fact that $N_{\a\b\c}=H_{\a\b\c}$), and the presence of these components provides another way of breaking to $SU(2)$. If the two ways of breaking to $SU(2)$, (\ref{eq:m4_reduction1}) and (\ref{eq:m4_reduction2}), agree, the structure group remains $SU(2)$, otherwise it reduces to the trivial group. Again, the covariant constancy of the components (\ref{eq:m4_reduction1}) and (\ref{eq:m4_reduction2}) follows from the general result  (\ref{eq:Hmixed_cov_constant}).

The $(4,0)$ and $(0,4)$ parts of (\ref{eq:dH_constraint}), together with the vanishing of $H_{ab4}$,  imply that the trace part of $H$ in the fourth direction vanishes:

\begin{equation}
\label{eq:trace4}
g^{a \overline{b}} H_{a \overline b 4}  =   g^{a \overline{b}} H_{a \overline b \overline{4}} = 0 \  .
\end{equation}

This also turns out to be the condition for the commutators between symmetries associated $v$ and $\hv$, and symmetries associated with the four-forms $L$ and $\hL$ to vanish.

As in the $SU(3)$ case, the symmetry due to $\tau_{I}$ needs to be introduced separately, as it is not generated from the original SU(4) special holonomy symmetries.  $\theta_{I}$ is proportional to $I_{ij} I_{kl} H^{jkl}$,  and it follows from  (\ref{eq:trace4}) it can have no component in the fourth direction. Therefore, the gradient of the dilaton field determines whether $\tau_{I}$ and $\omega_v$ are linearly independent.  The commutators (\ref{eq:vv_form2}) will then generate any additional invariant forms, the details of which depend on the direction in which the $\tau_{I}$ form is pointing.

\subsection{$G=Sp(k);\ n=4k$}
\label{subsec:symplectic}

Target spaces of this type could be called almost-complex HKT manifolds \cite{Howe:1996kj}. There is a set of Nijenhuis three-forms given by

\be
N^{rs}_{ijk}=\d^{rs}H_{ijk}-3 H_{lm[i}(I^r)^l{}_j (I^s)^m{}_{k]}\ ,
\la{3.29}
\ee

where $\{I^r\}$ is a set of three almost complex structures obeying
the algebra of the imaginary  quaternion units

\begin{equation}
\label{eq:quat_algebra}
I^r I^s = - \delta^{rs} + \epsilon^{rst} I^t \ .
\end{equation}

These forms do not vanish. One way of understanding their content is
to write a real vector  index $i=1\ldots 4k$ as a pair $i\rightarrow
a x$, where now $x=1\ldots 2k$ and $a=1,2$. The latter index is
acted on by the rigid $Sp(1)$ while the former is acted on by
$Sp(k)$. In this notation a general three-form $H$ can be written

\be
H_{ijk}\rightarrow H_{a x b y c z}= H_{(abc) [xyz]} + \e_{ac}
H'_{b yz,x} + \e_{bc} H'_{a zx,y}\ .
\la{3.30}
\ee

The $H$-tensor on the right has the indicated symmetries while
$H'_{axy,z}$ is antisymmetric on  the first two $Sp(k)$ indices with the
totally antisymmetric part being zero. In the Nijenhuis forms, this
part of $H$ drops out and so they are determined by $H_{a b c
xyz}$. In detail,

\be
N^{rs}_{a x b y c z}= (\s^r)_{( a}{}^{ d} (\s^s)_{ b}{}^{ e}
H_{ c) d e xyz}\ .
\la{3.31}
\ee

Adopting temporarily the notation $I^r= \{ I, J, K \} $, one can easily verify that there is only one independent covariantly constant object, since the covariant constancy of $N(I, I)$ implies, via the relations (\ref{eq:quat_algebra}), the covariant constancy of all the other Nijenhuis forms, $N(I,J)$, $N(J,J)$, and so on.

Starting from the $\theta$ forms,

\begin{equation}
\theta^{rs} \propto \star ( \star d \omega^{(r} \wedge \omega^{s)} )\ ,
\end{equation}

where $\omega^r$ is the two-form associated with $I^r$, one can define six covariantly constant vectors $\tau^{rs}$. However, as five of these involve $H_{abcxyz}$, only one of them is independent.


For $k=1$  the $N$ tensors vanish, and the manifold is four dimensional hyper-K\"{a}hler  with torsion. As $Sp(1) = SU(2)$, this case overlaps with the $m=1$ case of the previous subsection. If a $\tau$ form is present the structure group will reduce to the trivial group. For $k=2$ the situation is similar to the $m=4$ case in the last subsection, as $Sp(2) \subset SU(4)$. It follows that with $H_{abcxyz} \neq 0$ the structure group reduces to at most $SU(2)$.

\subsection{$G=Sp(k)\cdot Sp(1);\ n=4k$}

Manifolds of this type can be called almost QKT spaces
\cite{Howe:1997pn}. The only Nijenhuis form  is the seven-form
$\tN(L,L)$ arising from the four-form $L$. In dimension $8$ the dual
can only be proportional to $L_{ijkl} H^{jkl}$. Since this object is
not covariantly constant the constant of proportionality must be
zero, as can be checked explicitly. Alternatively one can show that
when $n=8$, $\tN(L,L)$ can be written as

\begin{equation}
\tN(L,L)_{ijklmrs} = 12 H_{[pqi} L^p_{jkl} L^q_{mrs]} \ \ \ ,
\end{equation}

which involves an antisymmetrisation over nine indices (an analogous
proof is for the vanishing  of the almost-complex Nijenhuis form in
four dimensions, by writing it as an object involving antisymmetrisation over five indices). For $k>2$ these arguments are not valid which
leads us to believe that $\tN(L,L)$ is not zero in general.

\subsection{$Spin(7)$ and $G_2$}

In $Spin(7)$ there is a particular self-dual four-form $\phi$ whose corresponding  Nijenhuis form $\tN(\phi,\phi)$
vanishes.  The Lee form $\tau_{\phi}$ is proportional to $\phi_{ijkl} H^{jkl}$, and if $\tau_{\phi}$ is non-vanishing the structure group reduces to $G_2$.  It generates an additional symmetry of the type (\ref{eq:v_symmetry}), which has a potentially non-vanishing commutator  with the $\phi$ symmetry. This is an $L$-type symmetry generated by a four-form. The algebra closes if this form vanishes, or if it is proportional to $\phi$. Otherwise the structure group is reduced further.


In $G_2$ there is a three-form $\varphi$ and its dual four-form $*\varphi$ and
the only non-zero Nijenhuis form is $\tN(*\varphi, *\varphi)$. This a seven-form
which is equal to the volume form multiplied by a constant times
$\varphi_{ijk} H^{ijk}$. It is easy to see that this function is constant
due to the structure of the $\gg_2$ Lie algebra.  The rest of the story is similar to the $Spin(7)$ case. The $\theta_{\varphi}$ form is proportional to $*\varphi_{ijkl} H^{jkl}$, and if $\tau_{\varphi}$ is non-zero the structure group reduces to $SU(3)$. The potentially non-vanishing Nijenhuis forms are generated in the commutator of the  of the related vector symmetry transformation with the $L$-type symmetries of $\varphi$ and $*\varphi$. These are a three- and a four-form, respectively. The algebra closes if these are zero or proportional to the original $G_2$ forms, and the structure group is further reduced otherwise.


\section{Supergravity and invariant forms}
\label{sec:sugra}

\subsection{Backgrounds with compact holonomy}
\label{sec:target_compact}

Consider a solution of the gravitino KSE for which the isotropy group of the $\nab^{(+)}$-parallel spinors is compact,
see eqn (\ref{ns1}).
To identify the ``Euclidean component'' of the spacetime,
we use the  results of \cite{het1, het2, het3a} on the solution of KSEs.
In particular, the gravitino KSE implies
that the Lorentzian spacetime admits  3, 4, 6
and 10 $\nab^{(+)}$-parallel, and so Killing, vectors, respectively. These are constructed as parallel spinor bi-linears.
If, in addition, one
assumes that the vector space spanned by these parallel vector
fields closes under Lie brackets, which is not always implied by the KSEs,
and the infinitesimal action can be integrated to a free $G$-group action, then the Lorentzian spacetime is a
principal bundle with fibre a Lorentzian Lie group $G$ and base
space $X_n$ for $n=7,6,4$ and $0$ dimensions, respectively. The
geometry of $X_n$ may depend on whether $G$ is abelian or
non-abelian. To extract the ``Euclidean component'', we shall take
$G$ to be {\it abelian}, the spacetime to be a metric product $G\times X_n$ and the fields to depend only on
the coordinates of $X_n$. Moreover,
we shall assume that $H$
has non-vanishing components only on $X_n$. Under these assumptions, $X_n$ is identified
as the ``Euclidean component'' of the Lorentzian spacetime.
With this definition, we exclude the parallelisable
manifolds with non-vanishing torsion, ie all non-abelian group manifold solutions of the
heterotic string\footnote{We could give another definition of the
``Euclidean component'' to include those but the above
definition will suffice for the applications we consider.}. Under these conditions,  $X_n$ are spaces, of dimension $n=7,6$ and $4$, admit a connection
$\nab^{(+)}$ with skew-symmetric torsion and  holonomy contained in $G_2$, $SU(3)$ and $SU(2)$, respectively.
In what
follows, we shall look at the conditions arising from dilatino KSE. In
particular, if one does not impose the dilatino KSE, then there are additional $\nab^{(+)}$-parallel forms on
$X_n$ to those constructed from $\nab^{(+)}$-parallel spinor bi-linears. This is  subject to imposing the Bianchi identity (\ref{3.24}),
$dH=0$ and the field equations of the heterotic
supergravity.
These new forms in turn lead to the reduction of the structure group to a
subgroup of the isotropy group  of $\nab^{(+)}$-parallel spinors. A similar reduction of the structure group for Lorentzian
manifolds has also been noticed in \cite{ het2} but its consequences on the geometry of spacetime were not extensively explored.

We shall show that each time that the holonomy of $\nab^{(+)}$ reduces, the gravitino KSE admits more parallel spinors. These in turn
give rise to more $\nab^{(+)}$-parallel forms which arise from  the conditions on the geometry imposed by
dilatino KSE. The new forms lead to further reduction of the holonomy of $\nab^{(+)}$. As a result,
the structure of $X_n$ reduces in various patterns. The geometry of $X_n$ at each stage can be determined using the results of \cite{het1, het2, het3a}.

\subsubsection{SU(2)}

The algebraically independent fundamental forms on $X_4$ are the Hermitian
forms $\omega_{\Lf}$ and $\omega_{\Qf}$ which are associated with almost
complex structures\footnote{The numerical subscript
attached to the almost complex structures denotes the dimension of the associated space.} $\Lf$ and $\Qf$ with $\Lf \Qf=-\Qf\Lf$, ie $X_4$ admits an
almost hyper-complex structure $(\Lf,\Qf,\Lf\Qf)$. The existence of a
compatible connection with skew-symmetric torsion requires that
\bea
N(\Lf)=N(\Qf)=0~,~~~i_{\Lf} d\omega_{\Lf}=i_{\Qf}d\omega_{\Qf}~.
\eea
Moreover, the torsion is given as
\bea
H=-i_{\Lf}d\omega_{\Lf}~.
\eea
Therefore $X_4$ is an HKT manifold. This is the full content of the gravitino KSE.

The dilatino KSE also requires\footnote{Our form conventions are
$\phi={1\over k!}\phi_{i_1\dots i_k} dx^{i_1}\wedge\dots\wedge dx^{i_k}$ and
$(\star \phi)_{i_1\dots i_{n-k}}={1\over (n-k)!} \epsilon^{j_1\dots j_k}{}_{i_{1}\dots i_{n-k}}\phi_{j_1\dots j_k}$.}
that
\bea
\tau_{\Lf}= \theta_{\Lf}-2d\Phi~,
\la{dsu2}
\eea
vanishes,
where
\bea
\theta_{\Lf}=-\star(\star d\omega_{\Lf}\wedge \omega_{\Lf})~,
\eea
is the Lee form of $\omega_L$.

Of course if $\tau_{\Lf}$ vanishes, the $SU(2)$ structure may
not reduce. However let us assume that
$\tau_{\Lf}\not=0$. In such a case  ${\rm
hol}(\nab^{(+)})\subseteq SU(2)$, $dH=0$,  the
identity (\ref{3.24})
and the field equations imply that
\bea
\nab^{(+)} \tau_{\Lf}=0~.
\eea
Complexifying the
typical fibre of $TX_4$ with respect to $\Lf$, $SU(2)$ acts on it with
the fundamental 2-dimensional complex representation. Since in this
representation, the isotropy group of a vector in $SU(2)$ is the
identity,  the structure reduces to $\{1\}$. Therefore $X_4$ is a
group manifold and so locally isometric to either $S^1\times S^3$ or $T^4$.

\subsubsection{SU(3)}

The fundamental forms of $X_6$ associated with an $SU(3)$ structure  are a Hermitian form $\omega_{\Rs}$ and a (3,0)-form $\chi$.
The 6-dimensional manifolds with $SU(3)$ structure and skew-symmetric torsion have been extensively investigated, see eg \cite{Strominger:1986uh,
hullb, ivanovgp, tseytpap, chiossi,gauntlett, lustb}.
For the manifold $X_6$ with an $SU(3)$ structure to admit a compatible connection
with  skew-symmetric torsion,
\bea
N(\Rs)_{ijk}=N(\Rs)_{[ijk]}~,~~~\theta_{\Rs}=\theta_{{\rm Re}\,\chi}~,
\la{transu3}
\eea
where $\theta_{\Rs}=-\star (\star d\omega_{\Rs} \wedge \omega_{\Rs})$ and
$\theta_{{\rm Re}\,\chi}=-{1\over2}\star (\star d{\rm Re}\,\chi\wedge {\rm Re}\,\chi)$, ie
the Nijenhuis tensor $N(\Rs)$ must be skew symmetric in all three indices. Moreover, the torsion 3-form is completely determined
in terms of the fundamental forms and the metric \cite{stefansu, het2} as
\bea
H=-i_{\Rs} d\omega_{\Rs}-2N(\Rs)=\star d\omega_{\Rs}-\star (\theta_{\Rs}\wedge \omega_{\Rs})+N(\Rs)~.
\la{hsu3}
\eea
This concludes the analysis of the gravitino KSE.

Turning to the dilatino KSE, one finds that
\bea
N(\Rs)~,~~~   \tau_{\Rs} ~,
\eea
must vanish. The gravitino and dilatino KSEs imply that
$X_6$ is a hermitian,  conformally balanced manifold with ${\rm hol}(\nab^{(+)})\subseteq (SU(3))$.  If in addition $X_6$ is compact
and $dH=0$, then it has been shown in \cite{ivanovgp} that it is  Calabi-Yau. A non-compact example can be found in
\cite{chamvol, tseytpap}.

\vskip 0.5cm

\leftline{\underline {Step 1}}

To investigate the pattern of
reduction of the structure group, we shall follow the analysis in
the $SU(2)$ case and assume that the conditions that arise in the
analysis of the dilatino KSE are not imposed.
Then   ${\rm hol}(\nab^{(+)})\subseteq SU(3)$, $dH=0$, the
identity (\ref{3.24}) and the field equations imply that
\bea
\nab^{(+)} N(\Rs)=0~,~~~\nab^{(+)} \tau_{\Rs}=0~.
\eea
The first condition does not necessarily  imply the reduction of the $SU(3)$
holonomy. Instead, $N(\Rs)$ is written as a linear combination
of ${\rm Re}\,\chi$ and ${\rm Im}\,\chi$
\bea
N(\Rs)=r_1\, {\rm Re}\,\chi+r_2\, {\rm Im}\,\chi~,
\label{nih}
\eea
where $r_1, r_2$ are real constants.  However, if
$\tau_{\Rs}\not=0$, the structure group reduces to
$SU(2)$. This is because $SU(3)$ acts with the fundamental
representation on the typical fibre of $TX_6$ complexified with
respect  to $R$, and the isotropy group in $SU(3)$ of a vector in
this representation is $SU(2)$.

\vskip 0.5cm
\leftline{\underline
{Step 2}}

Now suppose that $\tau_{\Rs}\not=0$ and so the
connection with skew-symmetric torsion of  $X_6$ has holonomy ${\rm
hol}(\nab^{(+)})\subseteq SU(2)$. In such case, the gravitino
KSE admits 4 additional $\nab^{(+)}$-parallel
spinors and so 8 in total\footnote{For the count the parallel spinors, we  view the backgrounds as solutions
of the gravitino KSE of heterotic supergravity.}. The fundamental forms that can be
constructed as $\nab^{(+)}$-parallel spinor bilinears are two
1-forms $e^a$, $a=5,6$, as well as the Hermitian forms $\omega_{\Lf}$
and $\omega_{\Qf}$ of endomorphisms $\Lf$ and $\Qf$, where
$\Lf^2=\Qf^2=-1_{4\times 4}$ and $\Lf\Qf=-\Qf\Lf$. Moreover
$i_a\omega_{\Lf}=i_a\omega_{\Qf}=0$, $a=1,2$, where $i_a$ denotes inner derivation with respect to the vector field
$e_a$ dual to 1-form $e^a$.
Compatibility with the $SU(2)$ structure requires that one of the 1-forms, say $e^5$,  must be
$\theta_{\Rs}-2d\Phi$ and the other is $e^6=\Rs e^5$.

To give the conditions that arise from the gravitino KSE, we adapt a frame $e=(e^a, e^i)$ on $X_6$, where
$e^a$ is identified with the first two parallel 1-forms and $e^i$ span the rest of the frame. Then we write
\bea
ds^2&=&\delta_{ab} e^a e^b+d\tilde s^2~,~~~i,j=1,2,3,4~,~~~a,b=5,6
\cr
H&=&{1\over2} H_{abi}\, e^a\wedge e^b\wedge e^i+{1\over2} H_{ija}\, e^i\wedge e^j\wedge e^a+\tilde H~,
\la{abijk}
\eea
where
\bea
d\tilde s^2:=\delta_{ij} e^i e^j~,~~~\tilde H&:=&{1\over 3!} H_{ijk} e^i\wedge e^j\wedge e^k~.
\eea
To analyze the gravitino KSE for the above background, we apply the
results of \cite{het2}, see appendix A, for the background $\bR^{3,1}\times X_6$
with ${\rm hol}(\nab^{(+)})\subseteq SU(2)$. In particular, the conditions that arise from the
gravitino KSE can be written as
\bea
(de^a)^{2,0+0,2}_{ij}&=&-{1\over2} \delta^{ab} (i_{\Lf}\nabla_b\omega_{\Lf})_{ij}~,~~~
\cr
(de^a)_{ij} \omega_{\Lf}^{ij}&=&\delta^{ab}(\nabla_b\omega_{\Qf})_{ij} \omega^{ij}_{\Pf}~,~~~\Pf=\Lf\Qf
\cr
N(\Lf)_{ijk}&=&N(\Qf)_{ijk}=0~,
\eea
where the (2,0) and (0,2) part has been taken with respect to $\Lf$,
and
\bea
i_a H=\delta_{ab}\, de^b~,~~~\tilde H=-i_{\Lf} \tilde d\omega_{\Lf}=-i_{\Qf} \tilde d\omega_{\Qf}~,
\eea
where $\tilde d$ is the restriction of the exterior derivative along the $e^i$ directions. Observe that the conditions
along the $e^i$ directions resemble those that we have found for the $SU(2)$ case above.

To proceed, we solve the dilatino KSE  for a background $\bR^{3,1}\times X_6$
with ${\rm hol}(\nab^{(+)})\subseteq SU(2)$ and 8 Killing spinors. In particular, we find that
\bea
\partial_a\Phi~,~~~(de^a)^{2,0+0,2}_{ij}~,~~~de^a_{ij}\,
\omega_{\Lf}^{ij}~,~~~[e_a, e_b]_i~,~~~
\tau_{\Lf}~,
\la{dsu22}
\eea
must vanish.
Again using
${\rm hol}(\nab^{(+)})\subseteq SU(2)$, the  identity
(\ref{3.24}), $dH=0$,  the field equations, and after some
calculation, one can show that the tensors with components given  in (\ref{dsu22}) are all $\nab^{(+)}$-parallel. The non-vanishing
of some of the tensors in (\ref{dsu22}) does not necessarily lead to the reduction of the holonomy. In particular,
if $\partial_a\Phi, de^a_{ij}\, \omega_{\Lf}^{ij}\not=0$, the $SU(2)$ holonomy does not reduce further.
Similarly, if $ (de^a)^{2,0+0,2}$, $(de^a)^{2,0+0,2}\not=0$, one  can set these tensors proportional to $\omega_{\Qf}$ and $\omega_{\Pf}$
and the holonomy does not reduce.
However if either $[e_a, e_b]_i\not=0$ or
$\tau_{\Lf}\not=0$, then $X_6$ admits an
additional linearly independent $\nab^{(+)}$-parallel 1-form. As in
the $SU(2)$ case investigated previously, the structure reduces to
$\{1\}$ and the spacetime is a group manifold.

It is worth remarking that  in the case that  $[e_a, e_b]_i=0$, $X_6$ admits two commuting $\nab^{(+)}$-parallel vector fields.
This is because the only 2-dimensional  metric Lie algebra with Euclidean signature is $\bR^2$.
If their action can be integrated
to a $T^2$ free group action, then $X_6$ is a principal $T^2$ bundle over a 4-dimensional manifold $B_4$. The geometry  on $B_4$ inherited
from that of $X_6$
depends on the properties of $de^a$. We have seen that $i_a\omega_{\Lf}=i_a\omega_{\Qf}=0$ but
\bea
({\cal L}_a\omega_{\Lf^r})_{ij}= 2 (de^a)^k{}_{[i} (\omega_{\Lf^r})_{j]k}~,~~~r=1,2,3~,
\eea
where $\Lf^1=\Lf$, $\Lf^2=\Qf$ and $\Lf^3=\Pf=\Lf \Qf$. Now if
\bea
de^a+\star de^a=2 f^{ar} \omega_{\Lf^r}~,
\la{selfdual}
\eea
then
\bea
{\cal L}_a\omega_{\Lf^r}=2 f_a{}^s \,\epsilon_s{}^{rt}\, \omega_{\Lf^t}~,
\la{lielie}
\eea
where $f$ is constant.
Thus in general, the hermitian forms are not invariant under the torus action and so they do not decent as hermitian forms on $B_4$.
There are two possibilities to consider. One is that $f=0$, ie
$ (de^a)^{2,0+0,2}=de^a_{ij}\, \omega_{\Lf}^{ij}=0$, then $\omega_{\Lf}$ and $\omega_{\Qf}$ decent as hermitian forms on $B_4$. Thus $B_4$
admits an $SU(2)$ structure compatible with a connection with
skew-symmetric torsion, ie $B_4$ is an HKT manifold. On the other hand, if $f\not=0$, the integrability condition of (\ref{lielie}) and $[{\cal L}_a, {\cal L}_b]=0$ imply that
a linear combination of the three hermitian forms $\omega_{\Lf^r}$ is invariant under the torus action, see \cite{het3a} for more details.
As a result, $B_4$
admits a $U(2)$ structure
compatible with a connection with
skew-symmetric torsion, ie $B_4$ is a KT manifold.

\vskip 0.5cm
\leftline{\underline
{Step 3}}

Now suppose that ${\rm
hol}(\nab^{(+)})\subseteq \{1\}$. In such a case, the spacetime is group manifold. The 6-dimensional
Euclidean signature metric group manifolds
are locally isometric to $SU(2)\times SU(2)$, $SU(2)\times T^3$ and $T^6$.
\subsubsection{G$_2$}
\label{geomg2}


Let  $\varphi$ and its dual $\star\varphi$ be the fundamental forms of a 7-dimensional manifold $X_7$ with a $G_2$ structure.
7-dimensional  manifolds with a $G_2$ structure have been investigated in \cite{g2str}.
Such  manifolds admit a compatible connection with skew-symmetric torsion \cite{stefang2}, iff
\bea
d\star\varphi=\theta\wedge\star\varphi~,
\eea
where $\theta$ is the Lee form of $\varphi$
\bea
\theta=-{1\over3}\star (\star d\varphi\wedge \varphi)~.
\eea
In such a case, the torsion is uniquely determined in terms of the fundamental forms  as
\bea
H=-{1\over6} (d\varphi, \star\varphi) \varphi+\star d\varphi-\star(\theta\wedge \varphi)~.
\eea
This is the full content of the gravitino KSE.

The dilatino KSE requires that
\bea
(d\varphi, \star\varphi)~,~~~\tau_{\varphi}  \ ,
\eea
must vanish.

\vskip 0.5cm
\leftline{\underline {Step 1}}

Suppose that the conditions which
arise from the dilatino Killing spinor are not imposed. In such a
case,  ${\rm hol}(\nab^{(+)})\subseteq G_2$, the identity (\ref{3.24}), $dH=0$ and the field
equations imply that
\bea
\nab^{(+)} (d\varphi, \star\varphi)=0~,~~~ \nab^{(+)} \tau_{\varphi}  =0~.
\eea
Of course if $(d\varphi, \star\varphi)\not=0$, the holonomy does not reduce because it is a scalar.
However, if $\tau_\varphi \not=0$, the structure group of $X_7$ reduces
from $G_2$ to $SU(3)$.

\vskip 0.5cm

\leftline{\underline {Step 2}}

Suppose now that ${\rm hol}(\nab^{(+)})\subseteq SU(3)$ on $X_7$. In such a case,
the gravitino KSE admits 2 additional parallel
spinors and so 4 in total. Moreover the parallel spinor bilinears on $X_7$ are a 1-form
$e^7$, a hermitian form $\omega_{\Rs}$,  $i_7 \omega_{\Rs}=0$ and $\Rs^2=-1_{6\times 6}$, and a (3,0)-form $\chi$, $i_7
\chi=0$,  where again $i_7$ denotes the inner derivation with respect to the vector field $e_7$ dual to the 1-form $e^7$.
Therefore the fundamental forms are $e^7$ and  those of
$SU(3)$ case above in the directions orthogonal to $e^7$.

Adapting a frame  $e=(e^i, e^7)$, $i=1, \dots, 6$,  the metric and 3-form  are written as
\bea
ds^2&=& (e^7)^2+d\tilde s^2~,~~~
H={1\over2} H_{7ij}\, e^7\wedge e^i\wedge e^j+\tilde H~,~~~
\cr
d\tilde s^2&=&\delta_{ij} e^i e^j~,~~~\tilde H={1\over3!} H_{ijk} e^i\wedge e^j\wedge e^k~.
\eea
Applying the results of \cite{het2} for the manifold $\bR^{2,1}\times X_7$, ${\rm hol}(\nab^{(+)})\subseteq SU(3)$,
where the metric and 3-form
on $X_7$ are given as above,
the gravitino KSE requires that
\bea
(de^7)^{2,0+0,2}_{ij}&=&-{1\over2} i_{\Rs} \nabla_7(\omega_{\Rs})_{ij}~,
\cr
(de^7)_{ij} \omega^{ij}_{\Rs}&=&{1\over 6} \nabla_7{\rm Re}\chi_{ijk} \,\, {\rm Im}\chi^{ijk}~.
\la{gg2su3}
\eea
These conditions are in addition to those in (\ref{transu3}) along the directions orthogonal to $e^7$. Moreover,
$i_7 H=de^7$ and $\tilde H$ is given as in (\ref{hsu3}).

Furthermore, the four $\nab^{(+)}$-parallel spinors also solve the  dilatino KSE provided that
\bea
\partial_7\Phi~,~~~N({\Rs})_{ijk}~~~~,~~~(de^7)_{ij} \omega^{ij}_{\Rs}~,~~~
(de^7)^{2,0+0,2}_{ij}~,~~~\tau_{\Rs}~,
\eea
vanish. The structure reduces to $SU(2)$  if either $(de^7)^{2,0+0,2}\not=0$ or $\tau_{{\Rs}}\not=0$.
On the other hand if $\partial_7\Phi$, $N({\Rs})$, and $(de^7)_{ij} (\omega_{\Rs})^{ij}$ do not vanish, there is no further reduction
of the $SU(3)$ structure
since these tensors are either scalars or can be chosen to be proportional to existing $\nab^{(+)}$-parallel forms as in (\ref{nih}).

Now suppose that the infinitesimal action of $e_7$ can be integrated to a $U(1)$ free group action. In such a case
$X_7$ is a principal $S^1$ fibration over a base space $B_6$. The geometric properties on $B_6$ depend on $de^7$.
As we have mentioned $i_7\omega_{\Rs}=i_7 \chi=0$ but
\bea
{\cal L}_7(\omega_{\Rs})_{ij}&= &2 (de^7)^k{}_{[i} (\omega_{\Rs})_{j]k}~,
\cr
{\cal L}_7\chi_{i_1i_2i_3}&=&-3(de^7)^k{}_{[i_1} \chi_{i_2i_3]k}~.
\eea
Thus the hermitian form $\omega_{\Rs}$ and the (3,0)-form $\chi$ may not decent on $B_6$. There are several cases to consider.
One is that $(de^7)^{2,0+0,2}=(de^7)_{ij} \omega^{ij}_{\Rs}=0$ which in turn implies that ${\cal L}_7\omega_{\Rs}={\cal L}_7\chi=0$.
Thus $B_6$  admits a $SU(3)$-structure  compatible connection with skew-symmetric torsion. Moreover,  $e^7$
is a principal bundle connection with curvature that obeys the Hermitian-Einstein  condition.
One can also take $(de^7)^{2,0+0,2}=0$ but $(de^7)_{ij} \omega^{ij}_{\Rs}\not=0$. In this case, $\omega_{\Rs}$ is invariant and
so descents to a hermitian form on $B_6$ but ${\cal L}_7\chi\not=0$. Therefore $B_6$ admits a $U(3)$ structure
compatible with a connection with skew-symmetric torsion but not an $SU(3)$ structure. Finally, if both
$(de^7)^{2,0+0,2}, (de^7)_{ij} \omega^{ij}_{\Rs}\not=0$, then ${\cal L}_7(\omega_{\Rs}), {\cal L}_7\chi\not=0$ and so $B_6$
admits just an $SO(6)$ structure.

\vskip 0.5cm
\leftline{\underline {Step 3}}

Suppose next that ${\rm hol}(\nab^{(+)})\subseteq SU(2)$. In such a case, $X_7$ admits 8 $\nab^{(+)}$-parallel spinors.  The
$\nab^{(+)}$-parallel forms constructed from the  spinor bilinears
are three 1-forms $e^a$, $a=5,6,7$, and hermitian forms $\omega_{\Lf}$ and $\omega_{\Qf}$, with $i_a\omega_{\Lf}=i_a \omega_{\Qf}=0$, associated
with the endomorphisms $\Lf, \Qf$, $\Lf^2=\Qf^2=-1_{4\times 4}$, $\Lf\Qf=-\Qf\Lf$. The analysis can proceed as for the reduction of the structure
from $SU(3)$ to $SU(2)$ in section 5.1.2 step 2.  The only difference is that there is an
additional parallel 1-form. The metric and 3-form can be written as
\bea
ds^2&=&\delta_{ab} e^a e^b+\delta_{ij} e^i e^j~,~~~i,j=1,2,3,4~,~~~a,b=5,6,7~,
\cr
H&=&{1\over3!} H_{abc} e^a\wedge e^b\wedge e^c+{1\over2} H_{abi}\, e^a\wedge e^b\wedge e^i+{1\over2} H_{ija}\, e^i\wedge e^j\wedge e^a+\tilde H~,
\eea

The rest of the formulae in section  5.1.2 step 2 on the conditions that arise from the gravitino
KSE still apply
provided that the indices  $a,b=5,6,7$.

Turning to the dilatino KSE, one finds that in addition to the tensors  in (\ref{dsu22})
\bea
H_{abc}
\eea
must vanish. The holonomy of $\nab^{(+)}$  further reduces provided some of the tensors in (\ref{dsu22})
do not vanish. The analysis is identical to that done for the $SU(3)$ case.

Assuming that the commutator the 3-vector field $e^a$ closes and their action can be integrated to a free group action,
$X_7$ is a either $T^3$ or a $SU(2)$ fibration over a 4-manifold $B_4$. The structure inherited on $B_4$ from $X_7$
depends on the conditions on $de^a$ and whether the Lie algebra of the vector fields is $\bR^3$ or $\mathfrak{su}(2)$.
Setting again $\Lf^1=\Lf$, $\Lf^2=\Qf$ and $\Lf^3=\Lf\Qf$, and the self-dual part of $de^a$, $a=5,6,7$,  as in (\ref{selfdual}), one again recovers
(\ref{lielie}) but now $a=5,6,7$. If the Lie algebra of the vector fields is abelian, $\bR^3$, then the analysis proceeds as in section 5.1.2 step 2.
$B_4$ can either have an $SU(2)$ or $U(2)$ structure  compatible with a connection
with skew-symmetric torsion depending on whether or not $f=0$. On the other hand if the Lie algebra of the vector fields is $\mathfrak{su}(2)$, there is an additional
possibility. This arises whenever $f$ is a non-degenerate $3\times 3$ matrix. The integrability condition
of  (\ref{lielie}) implies that it can always be  arranged such that $f$ is proportional to the identity. In such a case,
$B_4$ admits a $SU(2)\times SU(2)$ structure with anti-self-dual Weyl tensor, see also \cite{het3a}. The previous two cases arise whenever
$f$ is a degenerate $3\times 3$  or the zero matrix, respectively.

\vskip 0.5cm
\leftline{\underline
{Step 4}}

Now suppose that ${\rm
hol}(\nab^{(+)})\subseteq \{1\}$. In such a case, the spacetime is group manifold. The Euclidean signature metric group manifolds up to dimension 7
are locally isometric to $SU(2)\times SU(2)\times U(1)$, $SU(2)\times T^4$ and $T^7$.

\subsection{Backgrounds with non-compact holonomy}
\label{sec:target_noncompact}

The solution of the KSEs of  backgrounds for which ${\rm hol}(\nab^{(+)})$ is non-compact
and the reduction of the structure group of Lorentzian manifolds
have been investigated  in \cite{het1, het2}. These backgrounds always
admit a $\nab^{(+)}$-parallel null vector field.
To extract the ``Euclidean component'', we shall separate the light-cone directions from the rest.
For this,  we shall assume  that the spacetime is metrically $\bR^{1,1}\times X_8$,
the fields are independent from the
$\bR^{1,1}$ directions, and the 3-form $H$ has components only along the $X_8$ directions.
In such a case, the holonomy of the $\nab^{(+)}$ reduces to a subgroup of
\bea
Spin(7)~,~~~SU(4)~,~~~Sp(2)~,~~~\times^2Sp(1)~,~~~
Sp(1)~,~~~ U(1)~,~~~
\la{holnoncomp}
\eea
where we have excluded $\{1\}$ associated with the $\bR^8$ case in (\ref{noncomp}).
The number of $\nab^{(+)}$-parallel spinors of the holonomy groups (\ref{holnoncomp}) are as those in (\ref{noncomp}).
The first three groups in   (\ref{holnoncomp}) are in the
Berger list of holonomies for irreducible, simply connected Riemannian manifolds and act on the typical fibre
of $TX_8$ with the spinor, fundamental, and spinor representations,  respectively.  The other three holonomy
groups are new, and the way that  act on $TX_8$ will be described
later in detail, see also appendix B.

Taking the holonomy of $\nab^{(+)}$ as in (\ref{holnoncomp}), we shall again show that if
one does not impose the dilatino KSE, $X_8$
admits new parallel forms. These in turn reduce the holonomy of $\nab^{(+)}$ connection to  subgroups of (\ref{holnoncomp}).
This reduction leads to the existence of more parallel spinors on $X_8$ which again give new parallel forms associated with
the dilatino KSE. As a result the structure of $X_8$ reduces in patterns. One can determine the geometry of $X_8$ at each stage
by applying the results of \cite{het1,het2,het3a}.

\subsubsection{SU(4)}
\label{su4sec}


The fundamental forms of a manifold with an $SU(4)$-structure are a Hermitian form $\omega_I$, associated with an almost complex structure $\Ie$,
and a (4,0)-form $\psi$. In order a manifold with an $SU(4)$-structure to admit a compatible connection with skew-symmetric torsion
\bea
N(\Ie)_{ijk}=N(\Ie)_{[ijk]}~,~~~\theta_{\Ie}=
\theta_{{\rm Re}\,\psi}~,
\eea
where the Lee forms are
\bea
\theta_{\Ie}=-\star(\star d\omega_{\Ie}\wedge \omega_{\Ie})~,~~~\theta_{{\rm Re}\,\psi}=-{1\over4} \star(\star d{\rm Re}\,\psi\wedge {\rm Re}\,\psi)~,
\eea
and $N(\Ie)$  is a (3,0)- and (0,3)-form.
The torsion is completely determined in terms of the metric and the fundamental forms \cite{stefansu, het2} as
\bea
H=-i_{\Ie} d\omega_{\Ie}-2N(\Ie)=\star (d\omega_{\Ie}\wedge \omega_{\Ie})-{1\over2}\star (\theta_{\Ie}\wedge \omega_{\Ie}\wedge \omega_{\Ie})+N(\Ie)~.
\la{torsu4}
\eea

The dilatino KSE imposes the conditions that
\bea
N(\Ie)~,~~~\tau_{\Ie} ~,
\eea
must vanish.
The first condition implies that $X_8$ is a complex manifold   and the second
that $X_8$ is  conformally balanced.


\vskip 0.5cm

\leftline{\underline {Step 1}}

As in the previous cases, if the conditions that arise from the
dilatino Killing spinor are not imposed, they give rise to new
$\nab^{(+)}$-parallel forms on $X_8$. In particular, ${\rm
hol}(\nab^{(+)})\subseteq SU(4)$, the Bianchi indentity
(\ref{3.24}), $dH=0$ and the field equations imply that
\bea
\nab^{(+)} N(\Ie)=0~,~~~\nab^{(+)} \tau_{\Ie}=0~.
\eea
Now if the almost complex structure is not integrable dualising
$N(\Ie)$ with respect to ${\rm Re}\,\psi$, it gives rise to a
$\nab^{(+)}$-parallel 1-form $\tau$. Thus if either $N(\Ie)\not=0$ or
$\tau_{\Ie}\not=0$, the structure group reduces to
$SU(3)$. If both are non-zero and linearly independent, then the
structure group reduces to $SU(2)$.

\vskip 0.5cm

\leftline{\underline {Step 2}}

Now suppose that ${\rm hol}(\nab^{(+)})\subseteq SU(3)$. The spacetime admits two additional parallel spinors,
ie 4 in total.
In turn, $X_8$ admits two $\nab^{(+)}$-parallel 1-forms, $e^a$, $a=7,8$,
a Hermitian form $\omega_{\Rs}$ and a (3,0)-form $\chi$ such that $i_a\omega_{\Rs}=i_a\chi=0$.

The metric and $H$ can be written as in (\ref{abijk}) but now $a,b=7,8$ and $i,j,k=1,2,\dots,6$.
The conditions gravitino KSE imposes on $X_8$ are like those we have found
in the reduction of the structure group from $G_2$ to $SU(3)$, ie those stated in equations
(\ref{gg2su3}) and (\ref{transu3}). The only difference is that (\ref{gg2su3}) holds for two 1-forms
rather than one.

The dilatino KSE implies that
\bea
\partial_a\Phi~,~~~N(\Rs)_{ijk}~~~~,~~~(de^a)_{ij} \omega^{ij}_{\Rs}~,~~~
(de^a)^{2,0+0,2}_{ij}~,~~~[e_a, e_b]_i~,~~~\tau_{\Rs}~,
\eea
must vanish. On the other hand if one of the last three tensors do not vanish, they give rise to new $\nab^{(+)}$-parallel 1-forms on $X_8$
which reduce the structure further to a subgroup of $SU(2)$.

Assuming that $[e_a, e_b]_i=0$ and the action of the vector fields can be integrated to a $T^2$ free group action,
$X_8$ is a $T^2$ fibration over a 6-dimensional manifold $B_6$. The geometry inherited on $B_6$ from $X_8$ depends
on the properties of $de^a$. The analysis is similar to the one we have done at the end  section 5.1.3 step 2 for the reduction
from $G_2$ to $SU(3)$. The only difference is that in the latter case, the fibre direction is one rather than two.
Nevertheless the details remain the same.  $B_6$ admits an  $SU(3)$ or $U(3)$ structure compatible with a connection
with skew-symmetric torsion depending on whether
$(de^a)_{ij} \omega^{ij}_{\Rs}=(de^a)^{2,0+0,2}_{ij}=0$ or  $(de^a)_{ij} \omega^{ij}_{\Rs}\not=0, (de^a)^{2,0+0,2}_{ij}=0$, respectively.
Otherwise it admits a $SO(6)$ structure.

\vskip 0.5cm

\leftline{\underline {Step 3}}

Now suppose that  ${\rm hol}(\nab^{(+)})\subseteq SU(2)$.  The spacetime admits 8 $\nab^{(+)}$-parallel spinors in total.
In such a
case, $X_8$ admits four  $\nab^{(+)}$-parallel 1-forms, $e^a$, $a=5,6,7,8$, and a Hermitian forms $\omega_{\Lf}$ and $\omega_{\Qf}$, with
$\Lf^2=\Qf^2=-1_{4\times 4}$ and $\Lf\Qf=-\Qf\Lf$, such that $i_a\omega_{\Lf}=i_a\omega_{\Qf}=0$.  The solution
to the gravitino KSE and the analysis that
follows is similar to that we have explained for the reduction of
$G_2$  to $SU(2)$ structure, and so we shall not give details. The only difference is that in this case
there are four parallel 1-forms rather than three.
Moreover, if the action of the associated parallel vector fields
can be integrated to
a free group action, $X_8$ is a principal bundle over a 4-dimensional manifold $B_4$ with fibre either
$T^4$ or $SU(2)\times S^1$. An analysis similar to that done in section 5.1.3 step 3 reveals that for both fibres
$B_4$ admits an either $SU(2)$ or $U(2)$ structure compatible with a connection with skew-symmetric torsion.
If the fibre is $SU(2)\times S^1$, then there is an additional case that arises.  $B_4$ admits
an $SU(2)\times SU(2)$ structure with anti-self-dual Weyl tensor.

Furthermore, the structure can reduce to $\{1\}$. In such a case $X_8$ is a group manifold, and so locally isometric to $T^8$, $T^5\times SU(2)$,
$\times^2SU(2)\times T^2$ or $SU(3)$.

\subsubsection{Sp(2),~${\bf \times}^2$Sp(1),~Sp(1) and U(1)}


To describe the geometry, we begin with the $Sp(2)$ case. The
condition ${\rm hol}({\nab^{(+)}})\subseteq Sp(2)$ is equivalent to
requiring that $TX_8$ admits three endomorphisms $I,J$ and $K$ that
satisfy the algebra of imaginary unit quaternions, $\Ie^2=\Je^2=-1,
\Ke=\Ie\Je$. Moreover for each of the three almost complex structures, the
torsion $H$ can be written as (\ref{torsu4}) provided that the
associated Nijenhuis tensor is skew-symmetric in all indices. Since
if $I$ and $J$ are parallel, so is $K$, and since $H$ must be the
same for all almost complex structures, apart from the skew-symmetry
condition on the Nijenhuis tensors of $I$ and $J$, one also requires
that
\bea
i_{\Ie} d\omega_{\Ie}+2N(\Ie)=i_{\Je} d\omega_{\Je}+2N(\Je)~.
\eea
This is the
content of the gravitino KSE.

The dilatino KSE implies  that
\bea
N(\Ie)~,~~~N(\Je)~,~~~\tau_{\Ie}~,~~~\tau_{\Je}~,
\eea
must vanish.
So $X_8$ admits a hyper-complex structure, and so it is a conformally balanced HKT manifold.

To generalise the above discussion to ${\times}^2Sp(1)$, $Sp(1)$,  and $U(1)$, we shall follow \cite{het3} and observe that
in the $Sp(2)$ case the tangent bundle
of $TX_8$ is a ${\rm Cliff}(\bR^2)$ module, where ${\rm Cliff}(\bR^2)$ is taken with the negative definite inner product on $\bR^2$.
The basis $\{i,j\}$ of ${\rm Cliff}(\bR^2)$ are represented by $\{\Ie,\Je\}$, respectively,
while $\Ke$ corresponds to the even Clifford element $k=ij$. For the rest of the cases, $\times^2Sp(1), Sp(1)$ and $U(1)$,
$TX_8$ is a ${\rm Cliff}(\bR^n)$ module for $n=3,4$ and $5$, respectively. So progressively for each
new case, one has to introduce an additional almost complex structure on $TX_8$ which corresponds to an additional basis element of the
Clifford algebra which anti-commutes with all the previous ones. Thus in each case $TX_8$ admits the action of $n$ almost
complex structures $\Ie^r$, $r=1, 2, \dots, n$, which are all algebraically independent such that $(\Ie^r)^2=-1$, $\Ie^r \Ie^s=-\Ie^s \Ie^r$ for $r\not=s$.
Of course $TX_8$ admits the action of  almost complex and almost product structures which can be constructed by taking
products of the $n$ basis elements.
With these data, the conditions that arise from the gravitino KSE are
\bea
i_r d\omega_r+2N(\Ie^r)=i_s d\omega_s+2N(\Ie^s)~,~~~r\not=s
\eea
and that $N(\Ie^r)$, $r=1,\dots,n$, is skew-symmetric in all indices.

Similarly, the dilatino KSE implies that
\bea
N(\Ie^r)~,~~~\tau_{\Ie^r} ~,~~~r=1,\dots,n~,
\eea
must vanish. We have shown in appendix B that if ${\rm hol}(\nab^{(+)})\subseteq \times^2 Sp(1)$ and two commuting complex structures
are integrable, then $X_8$ factorizes to a product $X_8=X_4\times X'_4$ with $X_4$ and $X'_4$ each admitting an $Sp(1)$
structure compatible with a connection with skew-symmetric torsion. Under the same assumptions if the holonomy
of $X_8$ is a subgroup of $Sp(1)$ or $U(1)$, then $X_8$ is parallelisable.


Before we examine the additional parallel forms that arise for those backgrounds,
we shall first investigate the way that the structure groups $Sp(2)$, $\times^2Sp(1)$, $Sp(1)$ and $U(1)$ act on the typical
fibre of $TX_8$. Identifying the typical fibre of $TX_8$ with $\bH^2$,  $Sp(2)$, represented with $2\times 2$ matrices with entries
quaternions, acts on $\bH^2$ from the left. Then the action of $\Ie,\Je$ and $\Ke$ on the typical fibre can be identified with the
action  of $i,j$ and $k$, the quaternion basis, on $\bH^2$ from the right. Clearly, this action commutes with
that of $Sp(2)$, ie $\Ie,\Je$ and $\Ke$ are invariant under $Sp(2)$ as expected.

Next the structure group $\times^2Sp(1)$  can be identified with diagonal subgroup of $Sp(2)$. With this identification,
the action of $\times^2Sp(1)$ on $\bH^2$ commutes with the endomorphism $\Pi:~~x\oplus y\rightarrow x\oplus -y$.
Moreover $\Pi$ commutes with all $\Ie,\Je$ and $K$. So if $\{\Ie,\Je\}$ is chosen as the basis of ${\rm Cliff}(\bR^2)$ associated with the $Sp(2)$
case, then a basis for ${\rm Cliff}(\bR^3)$ is $\{\Ie^1, \Ie^2, \Ie^3\}=\{\Ie,\Je, \Pi \Ie \Je\}$.

Similarly, $Sp(1)$ structure group can be identified
with the diagonal subgroup of $\times^2 Sp(1)$. In such a case, the action of $Sp(1)$ on $\bH^2$ commutes with the endomorphism
$\Sigma: x\oplus y\rightarrow y\oplus x$. Moreover $\Sigma$ commutes with $\Ie^1, \Ie^2$ but anti-commutes with $\Ie^3$. The additional basis element
of ${\rm Cliff}(\bR^4)$ associated with $Sp(1)$ can be chosen as $\Ie^4=\Sigma \Ie\Je$.

It remains to investigate the action of  $U(1)$ structure group on $\bH^2$. If $Sp(1)$ is identified with the quaternions
of length one, then $U(1)$ is the subgroup of $Sp(1)$ spanned by the complex numbers of length one. It is clear then that
the action of $U(1)$ on $\bH^2$ commutes with the action $T$ of imaginary unit $i$ on $\bH^2$ acting from the left.
In addition, $T$ commutes with $\Ie^r$, $r=1,\dots, 4$, and $T^2=-1$. Using this the additional basis element
of ${\rm Cliff}(\bR^5)$ associated with the $U(1)$ holonomy group can be chosen as $\Ie^5=\Ie^1\Ie^2\Ie^3\Ie^4 T$. Observe that
$(\Ie^5)^2=-1$ and $\Ie^5$ anticommutes with the other four basis elements.

\vskip 0.5cm

\leftline{\underline{Step 1}}

Having established the action of the structure groups on the typical
fibre of $TX_8$, we can now investigate their reduction in the cases
that additional forms are $\nab^{(+)}$-parallel. To do this observe
that $X_8$ admits an $SU(4)$ structure with respect to each
almost complex structure $\Ie^r$. Suppose that the dilatino KSE is not satisfied. In such a case, the Nijenhuis tensor
$N(\Ie^r)$ may not vanish. Each such Nijenhuis tensor is
skew-symmetric in all three indices and (3,0)- and (0,3)- form with
respect to the associated almost complex structure $\Ie^r$. Dualising this with
the real part of (4,0)-form, one concludes that for each $N(\Ie^r)$
there is a 1-form on $X_8$. Moreover it turns out that $N(\Ie^r)$ is
$\nab^{(+)}$-parallel. For this, one again uses  ${\rm
hol}(\nab^{(+)})\subseteq K$, the identity
(\ref{3.24}), $dH=0$ and the field equations, where $K$ is $Sp(2)$, $\times^2 Sp(1)$, $Sp(1)$ or $U(1)$.
As a result for every non-vanishing Nijenhuis tensor
$N(\Ie^r)$, there is an associated $\nab^{(+)}$-parallel 1-form.
A similar calculation also reveals that $\tau_{\Ie^r}$
are also $\nab^{(+)}$-parallel. Of course the commutators of the
associated vector fields, if they are non-vanishing, they are also
$\nab^{(+)}$-parallel. If one or more such forms are non-vanishing,
the structure groups reduce. In particular for $Sp(2)$, if there are one or more linearly
independent
$\nab^{(+)}$-parallel 1-forms $\tau$, the structure
group reduces to either $Sp(1)=SU(2)$ or $\{1\}$.  The reduction pattern for $\times^2
Sp(1)$ is similar. In the $Sp(1)$ case, if there is parallel 1-form,
the structure group reduces to the identity. A similar result holds
for the $U(1)$ case.

\vskip 0.5cm
\leftline{\underline{Step 2}}

Now suppose that  ${\rm hol}(\nab^{(+)})\subseteq SU(2)$.
Observe that the action of this $SU(2)$ is different from that of $Sp(1)$, which is associated with
5 parallel spinors, on
the typical fibre of
$TX_8$.  In particular, the spacetime $\bR^{1,1}\times X_8$ with ${\rm hol}(\nab^{(+)})\subseteq SU(2)$ admits 8
parallel spinors. These give rise to four
$\nab^{(+)}$-parallel forms, $e^5, e^6, e^7$ and $e^8$, and Hermitian
forms $\omega_{\Lf}$ and $\omega_{\Qf}$ on $X_8$ obeying the algebraic relations
which have already been stated in the investigation
of the reduction of the  $SU(4)$-structure to $SU(2)$. The details of the analysis
are  similar to those of the $SU(4)$ case and so
we shall not expand further
here. This similarity extends whenever   the structure groups reduce to $\{1\}$.

\subsubsection{Spin(7)}


Eight-dimensional manifolds with a $Spin(7)$-structure
have been investigated in \cite{fred}. It is known that any 8-dimensional manifold with $Spin(7)$-structure admits a compatible connection
with skew-symmetric torsion \cite{ivanovspin7}.   So the gravitino KSE
can be solved for every 8-dimensional manifold which admits a $Spin(7)$ structure. Moreover, the torsion is completely determined in terms of
the metric and fundamental self-dual 4-form $\phi$ as
\bea
H=-\star d\phi+\star(\theta \wedge \phi)~,
\eea
where $\theta_\phi=-{1\over6}\star (\star d\phi\wedge \phi)$ is the Lee 1-form.
The torsion 3-form is not always closed.

The dilatino KSE requires that
\bea
\tau_{\phi} ~,
\eea
must vanish.

\leftline{\underline{Step 1}}

Now suppose that we have a solution of the gravitino KSE only, and so ${\rm hol}(\nab^{(+)})\subseteq Spin(7)$. Then
using, ${\rm hol}(\nab^{(+)})\subseteq Spin(7)$,  $dH=0$,  the
identity (\ref{3.24}) and the field equations, one finds
that
\bea
\nab^{(+)}  \tau_\phi  =0~.
\eea
It is clear
that if the dilatino KSE is satisfied, ie
$\tau_\phi=0$, there is no reduction of the $Spin(7)$ structure of
$X_8$. However if the dilatino KSE is not
satisfied, and so $\tau_\phi \not=0$,  the holonomy of
$\nab^{(+)}$ reduces to a subgroup of $Spin(7)$. To identify this
subgroup,  note that $Spin(7)$ acts with the spinor representation
on the typical fibre of   $TX_8$, and there is  one type of a
non-trivial orbit with isotropy group $G_2$. Therefore the holonomy
of $\nab^{(+)}$ and so the structure group of $X_8$ reduces to a
subgroup of $G_2$, ie ${\rm hol}(\nab^{(+)})\subseteq G_2$.

\vskip 0.5cm

\leftline{\underline{Step 2}}

Now if ${\rm hol}(\nab^{(+)})\subseteq G_2$, the gravitino KSE of $X_8$ admits an additional $\nab^{(+)}$-parallel
spinor. In such case, the  $\nab^{(+)}$-parallel forms bilinears
on $X_8$ are an 1-form $e^8$, and the fundamental $G_2$ 3- and
4-forms $\varphi$ and $\star_{{}_7}\varphi$, respectively, such that $i_8\varphi=i_8\star_{{}_7}\varphi=0$, where the subscripted star
denotes the Hodge duality operation in directions orthogonal to $e^8$.  Since the reduction to $G_2$ has been mediated
by the non-vanishing $\tau_\phi$ 1-form,  $e^8=\tau_\phi$.

Adapting a local frame as $(e^i, e^8)$, the metric  and 3-form are written as
\bea
ds^2&=&(e^8)^2+\delta_{ij} e^i e^j~,~~~i,j=1,\dots,7~,
\cr
H&=&{1\over2}H_{8ij}\, e^8\wedge e^i\wedge e^j+{1\over 3!} H_{ijk}\, e^i\wedge e^j\wedge e^k~.
\la{ghspin7}
\eea
Applying the results of \cite{het2} on the manifold $\bR^{1,1}\times X_8$, where the metric and torsion on $X_8$
are given as above, the gravitino KSE requires the conditions
\bea
( d e^8)_{ij}|_{\bf 7}={1\over6} \nabla_8\varphi_{mn[i} \varphi^{mn}{}_{j]}~,
\eea
and $i_8H=de^8$, in addition to those stated in section \ref{geomg2} for the directions orthogonal to $e^8$.

The dilatino KSE for the (\ref{ghspin7}) background implies that
\bea
\partial_8\Phi~,~~~\tilde de^8|_{\bf 7}~,~~~\tau_\varphi~,
\eea
must vanish, where $\tilde d$ is the exterior derivative restricted to directions orthogonal to $e^8$.
If $\partial_8\Phi\not=0$, the $G_2$ does not reduce further. However if either $\tilde de^8|_{\bf 7}\not=0$ or $\tau_\varphi \not=0$,
the structure reduces to $SU(3)$. If both do not vanish and are linearly independent,
then the structure reduces to $SU(2)$.

\vskip 0.5cm

\leftline{\underline{Step 3}}

It remains to investigate the further reduction of $SU(3)$- and
$SU(2)$-structures. The analysis is similar to that that we have already described in the $SU(4)$ case in section
\ref{su4sec},
so  we shall not pursue this further.


\section{Conclusions}
\label{sec:conclusion}


In this paper we have seen that when $H$-flux is turned on many of the
special holonomy algebras are potentially deformed by currents that are associated
with generalised Nijenhuis forms. The general analysis from the worldsheet perspective is rather complicated in arbitrary dimension in the absence of integrability, but as demonstrated in this paper it is tractable in lower dimensions.   Fortunately these are also the physically relevant cases. We were also able to analyse the implications for the worldsheet algebras due to various covariantly constant one-forms whose presence is justified by imposing conformal invariance, i.e. assuming the stringy equations of motion.


The general situation is that there are further reductions in the structure group that was initially taken to be determined by the presence of special holonomy forms. In the second part of the paper, adopting a spacetime point of view, we classified all possible reductions of this type using the Killing Spinor Equations.


{\bf Acknowledgements:}

GP is
partially supported
by the EPSRC grant EP/F069774/1 and the STFC rolling grant ST/G000/395/1. VS was supported by the DFG (German Science Foundation) and the University of Hamburg, as well as, in part, by a NWO VICI grant.  VS is also grateful for the support of Per Sundell and the University of Mons (F.R.S.-FNRS Ulysse Incentive Grant for Mobility in Scientific Research).

\appendix

\section{Notation and Conventions}


$(1,1)$ superspace has coordinates $z=(x^{++},x^{--},\th^+,\th^-)$.
$D_+$ and $D_-$ are the usual flat superspace  covariant derivatives
which obey the relations

\begin{equation}
D_+^2 = i\del_{++};\qquad D_-^2 = i\del_{--};\qquad \{D_+,D_-\}= 0 \
. \label{2.3}
\end{equation}

We use the convention that $\del_{++} x^{++}=1$.

The action for a $(1,1)$-supersymmetric  sigma model without boundary is

\begin{equation}
S=\int\, dz\, (g_{ij}+b_{ij}) D_+ X^i D_- X^j\ , \label{2.1}
\end{equation}

where $dz:=d^2x\,D_+ D_-$. The action \eq{2.1} is invariant under
superconformal transformations which act  independently on the left
(+) and right (-) light-cone sectors. In the left sector, the
supercurrent is the energy-momentum tensor

\begin{equation}
T_{+3}:=g_{ij} \del_{++}X^i D_+X^j -\frac{i}{6}H_{ijk} D_+ X^{ijk}\
, \label{2.8}
\end{equation}

where we have introduced the abbreviation

\begin{equation}
D_+X^{ijk}:=D_+ X^i D_+ X^j D_+ X^k \ .
\label{2.9}
\end{equation}

The current is conserved in the sense that $D_-T_{+3}=0$ on-shell.
Similarly, there is a conserved  energy-momentum tensor $T_{-3}$ in
the right sector.

The action of the $(1,0)$ model is given by

\begin{equation}
\label{eq:heterotic_sigma}
S = \int d x^{++} dx^{--}  d \theta^+ (g_{ij} + b_{ij} ) D_+ X^i \partial_{--} X^j  \ .
\end{equation}

In the left sector we have a supercurrent as in the $(1,1)$ model, (\ref{2.8}), but in the right sector the conserved current is just:

\begin{equation}
T_{-3} = g_{ij} \partial_{++} X^i \partial_{++} X^j \ .
\end{equation}

Let $L$ be a vector-valued $l$-form such that the $l+1$-form
obtained by lowering the vector index  (taken to be in the first
slot) is covariantly constant with respect to $\nab^\pl$; this form
will also be denoted $L$. (It should be clear from the context which
is meant). The actions  of the $(1,1)$ and the $(1,0)$ models are both  are invariant under the transformation

\begin{equation}
\d_L X^i=a_L L^i{}_L D_+ X^L  \ , \label{2.10B}
\end{equation}

where the parameter $a_L$ has Lorentz weight $-l$, Grassmann parity
$(-1)^l$ and is chiral, $D_- a_L=0$.  The multi-index $L$ denotes
$l$ antisymmetrised indices, $L:=[l_1\ldots l_l]$. We shall use the
notation $L_2$ to denote antisymmetrisation over the $l-1$ indices
beginning with $l_2$, and so on. Analogous $L$-type symmetries exist in the right sector only in the case of the  $(1,1)$ model.

\section{Factorization of geometries with special holonomy}

Here we shall show that if  ${\rm hol}(\nabla^{(+)})\subseteq \times^2Sp(1)$ and two commuting  complex structures
are integrable, then $X_8$ locally factorizes as  $X_8=X_4\times X'_4$, where $X_4$ and $X'_4$  admit an $Sp(1)$
structure compatible with a connection with skew-symmetric torsion.
Moreover under the same assumption, if  $\nabla^{(+)}$ has
holonomy either $Sp(1)\subset \times^2Sp(1)$ or $U(1)\subset
Sp(1)\subset \times^2Sp(1)$, then $X_8$ is parallelisable.

{}First let us begin with $\times^2Sp(1)$. This follows from the
more general result that if the holonomy of $\nabla^{(+)}$ is
$U(n)\times U(m)$ and the associated complex and product structures
are integrable, then $X_{2n+2m}$ locally factorizes as $X_{2n+2m}=X_{2n}\times X_{2m}$, where where $X_{2n}$ and $X_{2m}$  admit a $U(n)$
and a $U(m)$
structure compatible with a connection with skew-symmetric torsion, respectively. Indeed if $\nabla^{(+)}$
has holonomy $U(n)\times U(m)$, it admits two commuting complex
structures $I,J$, $I^2=J^2=-1$ and $IJ=JI$. $\Pi=IJ$ is a product
structure and it is integrable provided that both $I$ and $J$ are
integrable. In such a case, there is an atlas on $X_{2n+2m}$ such that
\bea
I=(i\delta^\a{}_\b, i\delta^\mu{}_\nu, -i\delta^{\bar\a}{}_{\bar\b}, -i\delta^{\bar\mu}{}_{\bar \nu})~,
\cr
J=(i\delta^\a{}_\b, -i\delta^\mu{}_\nu, -i\delta^{\bar\a}{}_{\bar\b}, i\delta^{\bar\mu}{}_{\bar \nu})~,
\eea
where $(z^\a, w^\mu)$, $\a=1,\dots, n$, $\mu=1,\dots,m$,  are holomorphic coordinates and the transition functions are holomorphic respecting the splitting.
The integrability of the complex structures implies that the non-vanishing components of $H$ are
\bea
H_{\a\b\bar\gamma}~,~~~H_{\mu\bar\nu\alpha}~,~~~H_{\a\bar\b\mu}~,~~~H_{\mu\nu\bar\rho}~,
\eea
and their complex conjugates. Since the metric is hermitian with respect to both complex structures,
the non-vanishing components of the
metric are
\bea
g=(g_{\a\bar\b}, g_{\mu\bar\nu})~.
\eea
So far the components of the torsion $H$ and the metric depend on all coordinates.

Since $I$ and $J$ are $\nabla^{(+)}$-covariantly constant,  $H$ is determined in terms of both the complex structures $I$ and $J$ leading
to the condition
\bea
H=-i_I d\omega_I=-i_Jd\omega_J~.
\eea
Evaluating $H_{\mu\bar\nu\alpha}$ using both the $I$ and $J$ complex structures, one finds that $H_{\mu\bar\nu\alpha}=-3\partial_\a g_{\mu\bar\nu}$
and $H_{\mu\bar\nu\alpha}=3\partial_\a g_{\mu\bar\nu}$, respectively. Therefore consistency requires that
\bea
\partial_\a g_{\mu\bar\nu}=0~,~~~H_{\mu\bar\nu\alpha}=0~,
\eea
and similarly
\bea
\partial_\mu g_{\a\bar\b}=0~,~~~H_{\a\bar\b\mu}=0~.
\eea
As a result, $X_{2n+2m}$ is metrically locally a product, $X_{2n+2m}=X_{2n}\times X_{2m}$. $X_{2n}$ is a hermitian manifold with complex structure
$I_1=(i\delta^\a{}_\b, -i\delta^{\bar\a}{}_{\bar\b})$ and metric $g_1=(g_{\a\bar\b})$ and so with torsion $H_1=(H_{\a\b\bar\gamma},
H_{\bar\a\bar\b\gamma})$, where all components of $g_1$ and $H_1$ depend only on the coordinates $(z^\a, z^{\bar\a})$. Similarly,
$X_{2m}$ is a hermitian manifold with complex structure
$I_2=(i\delta^\mu{}_\nu, -i\delta^{\bar\mu}{}_{\bar\nu})$ and metric $g_2=(g_{\mu\bar\nu})$ and torsion $H_2=(H_{\mu\nu\bar\rho},
H_{\bar\mu\bar\nu\rho})$, where all components of $g_2$ and $H_2$ depend only on the coordinates $(w^\mu, w^{\bar\mu})$.

A consequence of the result above is that if $X_8$ has holonomy $\times^2Sp(1)\subset \times^2U(2)$, then $X_8$ is the
product of two 4-dimensional HKT manifolds. This proves the statement for holonomy $\times^2Sp(1)$.

To investigate the holonomy $Sp(1)\subset \times^2Sp(1)$ and $U(1)\subset Sp(1)\subset \times^2Sp(1)$ manifolds, note that in both these
cases there is an (almost) complex  structure $K$ on $\cM$ such that $\nabla^{(+)} K=0$ and
\bea
\omega_K=K_{\a\mu} dz^\a\wedge dw^\mu+K_{\bar\a\bar\mu} dz^{\bar\a}\wedge dw^{\bar\mu}~.
\eea
The integrability condition of $\nabla^{(+)} K=0$ gives
\bea
R^{(+)}_{mn, ij} K^m{}_p K^n{}_q= R^{(+)}_{pq, ij}~.
\eea
Using that $X_8$ is a product and so the Riemann tensor factorizes, one can easily show that $R^{(+)}=0$. Thus $X_8$ has
trivial holonomy and  so it is parallelisable. Moreover if $dH=0$, then $X_8$ is a group manifold.


\end{document}